\documentclass[12pt]{article}
\pdfoutput=1
\usepackage{graphicx,epstopdf,amssymb,amsfonts,amsmath,amsthm,array,
mathrsfs,amscd}
\usepackage{hyperref}
\usepackage[vcentermath]{youngtab}
\newcommand{\pbs}[1]{\let\temp=\\#1\let\\=\temp}
\numberwithin{equation}{section}
\def\be{\begin{equation}}\def\ee{\end{equation}}
\def\cvp{\raise 2pt\hbox{,}} 
 \def\tr{\mathop{\rm tr}\nolimits}
\def\rk{\mathop{\text{rk}}\nolimits}
\def\im{\mathop{\rm Im}\nolimits}
\def\re{\mathop{\rm Re}\nolimits}  

 \def\d{{\rm d}}\def\nn{{\cal
N}}

\def\gs{g_{\text s}}\def\ls{\ell_{\text s}}
\def\e{\epsilon}\def\a{\alpha}\def\b{\beta}
\def\cn{c_{i_{1}\cdots i_{n}}}\def\gn{\gamma^{i}_{i_{1}\cdots i_{n}}}
\def\an{\alpha^{i}_{i_{1}\cdots i_{n}}}
\def\bn{\beta^{i}_{i_{1}\cdots i_{n}}}
\def\gn{\gamma^{i}_{i_{1}\cdots i_{n}}}
\def\GD{G_{\text{D-geom}}}\def\GDL{\mathfrak G_{\text{D-geom}}}
\def\Sn{\text{S}_{n}}\def\Zn{\mathbb Z_{n}}
\def\Sno{\text{S}_{n+1}}
\def\jZn{j_{\Zn}}
\def\iin{i_{1}\cdots i_{n}}
\def\CSn{\mathbb C[\Sn]}
\def\Str{\mathop{\text{Str}}\nolimits}
\theoremstyle{plain}

\theoremstyle{definition}
\theoremstyle{remark}

\def\plb#1#2#3{{\it Phys.\ Lett.\ }{\bf B #1} (#2) #3}
\def\npb#1#2#3{{\it Nucl.\ Phys.\ }{\bf B #1} (#2) #3}
\def\npps#1#2#3{{\it Nucl.\ Phys.\ Proc.\ Suppl.\ }{\bf #1} (#2) #3}

\def\jhep#1#2#3{{\it J. High Energy Phys.\ }{\bf #1} (#2) #3}
\def\prd#1#2#3{{\it Phys.\ Rev.\ }{\bf D #1} (#2) #3}

\def\atmp#1#2#3{{\it Adv.\ Theor.\ Math.\ Phys.\ }{\bf #1} (#2) #3}

\def\ap#1#2#3{{\it Ann.\ of Phys.\ }{\bf #1} (#2) #3}

\def\fortphys#1#2#3{{\it Fortsch.\ Phys.\ }{\bf #1} (#2) #3}
\def\imath#1#2#3{{\it Invent math }{\bf #1} (#2) #3}

\begin{document}
%
%
{\pagestyle{empty}
\parskip 0in
\

\vfill
\begin{center}



{\LARGE On Matrix Geometry and Effective Actions}

\vspace{0.4in}

Frank F{\scshape errari}
\\
\medskip
{\it Service de Physique Th\'eorique et Math\'ematique\\
Universit\'e Libre de Bruxelles and International Solvay Institutes\\
Campus de la Plaine, CP 231, B-1050 Bruxelles, Belgique}
\smallskip
{\tt frank.ferrari@ulb.ac.be}
\end{center}
\vfill\noindent

We provide an elementary systematic discussion of single-trace matrix actions and of the group of matrix reparameterization that acts on them. The action of this group yields a generalized notion of gauge invariance which encompasses ordinary diffeomorphism and gauge invariances. We apply the formalism to non-abelian D-brane actions in arbitrary supergravity backgrounds, providing in particular explicit checks of the consistency of Myers' formulas with supergravity gauge invariances.  
We also draw interesting consequences for emergent space models based on the study of matrix effective actions. For example, in the case of the 
$\text{AdS}_{5}\times\text{S}^{5}$ background, we explain how the standard tensor transformation laws of the supergravity fields under ordinary diffeomorphisms emerge from the D-instanton effective action in this background.

\vfill

\medskip
%
\begin{flushleft}
\today
\end{flushleft}
\newpage\pagestyle{plain}
\baselineskip 16pt
\setcounter{footnote}{0}

}

\section{\label{s1} Introduction}
%


Recently, the author has proposed a general strategy to build calculable models of emergent space, based on a slightly modified version of the usual AdS/CFT correspondence \cite{fer1}: instead of considering the scattering of closed string modes off a large number $N$ of background branes, which yields ordinary gauge invariant correlators from the point of view of the worldvolume theory, one considers the scattering of a fixed number $k$ of probe D-branes. The pre-geometric, microscopic theory on the probe branes in the presence of the background branes can be solved at large $N$ \cite{solN}. The result is an effective action $S(X)$, expressed in terms of Hermitian matrix variables $X^{i}$ of size $k\times k$, whose fluctuations are suppressed at large $N$. These matrices can be interpreted as being the emergent matrix target space coordinates for the probe branes embedded in the ten-dimensional supergravity background sourced by the background branes. The action $S(X)$ should then correspond to the full non-abelian action for the probe branes in this emergent background. By studying the expansion of $S(X)$ around diagonal configurations,
\be\label{Xexp} X^{i}=x^{i}\mathbb I + \e^{i}\, ,\ee
the full supergravity background can actually be read off unambiguously, by comparing with the known form of the non-abelian action for D-branes in a general background \cite{Myers}. 

For example, if the probe branes are D-instantons in the type IIB theory, the large $N$ action $S(X)$ is a single-trace function of the matrices $X$ and its expansion takes the general form
\be\label{Sexp} S(X) =\sum_{n\geq 0}\frac{1}{n!}c_{i_{1}\cdots i_{n}}(x)\tr
\e^{i_{1}}\cdots\e^{i_{n}}\, .\ee
The cyclicity of the trace implies that the coefficients $\cn$ satisfy the cyclicity condition $c_{i_{1}\cdots i_{n}}=c_{i_{n}i_{1}\cdots i_{n-1}}$.
On the other hand, from Myers' formulas \cite{Myers}, we can relate certain combinations of the coefficients $\cn$ to the supergravity fields \cite{fer1},
\begin{align}
\label{c0} &c =-2i\pi(ie^{-\phi}-C_{0}) = - 2i\pi\tau\, ,\\
\label{c3} &c_{[ijk]}  = -\frac{12\pi}{\ls^{2}}\partial_{[i}( \tau B - C_{2})_{jk]}\, ,\\
\label{c4} &c_{[ij][kl]} = -\frac{18\pi}{\ls^{4}}e^{-\phi} 
\bigl(G_{ik}G_{jl}-G_{il}G_{jk}\bigr)\, ,\\
\label{c5}
&c_{[ijklm]}  = -\frac{120 i \pi}{\ls^{4}}
\partial_{[i}\bigl( C_{4} + C_{2}\wedge B - \frac{1}{2}\tau B\wedge B
\bigr)_{jklm]}\, ,
\end{align}
where $\phi$, $B$, $G$, $C_{0}$, $C_{2}$ and $C_{4}$ are the dilaton, the Kalb-Ramond two-form, the string frame metric and the Ramond-Ramond form potentials respectively. By matching the coefficients $\cn$, computed from the large $N$ solution of the microscopic model for the D-instanton in the presence of $N$ D3-branes, with \eqref{c0}, \eqref{c3}, \eqref{c4} and \eqref{c5}, the full 
$\text{AdS}_{5}\times\text{S}^{5}$ background was derived in \cite{fer1}. For instance, the coefficient $c_{[ijklm]}$ allows to find the non-trivial five-form field strength $F_{5}=\d C_{4}$ which, quite remarkably from the point of view of the microscopic model, turns out to be self-dual and normalized consistently with the Dirac-quantized D3-brane charge in string theory. 

The emergent geometry point of view that we have just outlined raises many questions on the general properties of the matrix action $S(X)$. The aim of the present technical note is to address some of these questions, bringing a better understanding of general properties of non-abelian D-brane actions and
providing interesting consistency checks of the approach introduced in \cite{fer1}. 

A basic set of questions is related to the general form of the expansion \eqref{Sexp}. The coefficients given by the equations 
\eqref{c0}--\eqref{c5} have rather non-trivial and surprising properties. For example, the coefficients $c_{[ijk]}$ and $c_{[ijklm]}$ automatically satisfy the constraints
\be\label{anti3and5} \partial_{[i}c_{jkl]}=0\, ,\quad
\partial_{[i}c_{jklm]} = 0\, .\ee
One may assume that these properties are accidents of the leading $\ls^{2}=2\pi\alpha'$ approximation used in Myers'. Indeed, formulas \eqref{c0}, \eqref{c3}, \eqref{c4} and \eqref{c5} are corrected, in general, by higher derivarive terms generated by the small $\ls^{2}$ expansion of appropriate disk string diagrams. When evaluated on highly supersymmetric backgrounds like the $\text{AdS}_{5}\times\text{S}^{5}$ background studied in \cite{fer1}, these corrections do vanish, but they will not on an arbitrary background. It thus came as a surprise to the author when calculations made in rather complicated examples \cite{fmr,fm,fr}, including cases with no supersymmetry at all, were found to yield results consistent with \eqref{anti3and5}. 

This led us to study the most general form of the Taylor expansion of a single-trace matrix function. Even though this is a rather elementary question with a quite useful solution, yielding approximation-independent constraints on the hard-to-compute (see e.g.\ \cite{Myers,Sevrin} and references therein) single-trace effective potential in any matrix theory, we have not been able to find a systematic discussion in the literature and we thus provide one in Section 2. 
In particular, we show that the conditions \eqref{anti3and5} and their generalizations to higher orders must always be valid, for any single-trace function $S(X)$. This has an interesting consequence: the formulas \eqref{c3} and \eqref{c5} can be used to define in a very natural way the supergravity $p$-form fields to all orders in $\ls^{2}$, and even at finite $\ls^{2}$. We also show that many combinations of the coefficients 
$\cn$ that are not listed in \eqref{c0}--\eqref{c5} can actually be expressed in terms of \eqref{c0}--\eqref{c5}, consistently with the rather complicated-looking form of Myers' formulas for these coefficients
(see for example the equation (5.8) in \cite{fer1}).

Another set of interesting questions is related to the physical content of the matrix actions $S(X)$. In other words, what are the natural gauge invariances of a general matrix action? 

The most traditional point of view on this problem is to start from the known diffeomorphism and $p$-form gauge invariances and try to check and/or impose them on the non-abelian D-brane actions \cite{gc,ginv,gcinv}. This is quite non-trivial. For example, in \cite{gc} it is clearly explained that the ordinary group of diffeomorphisms cannot act consistently on non-commuting matrix coordinates $X^{i}$, a result that will follow straightforwardly from our analysis in Section 3. The invariance under the $p$-form gauge symmetries is also quite involved \cite{ginv,gcinv}, since it is the $p$-form
potentials, not the gauge invariant field strengths, that enter into the non-abelian D-brane actions. As was shown in \cite{ginv}, by focusing on the Chern-Simons part of the action, consistency requires that the matrix coordinates must transform non-trivially under the $p$-form gauge symmetries, a rather surprising result rooted in the non-commutative nature of the space-time coordinates. We provide in Sections 4 and 5 explicit checks of the consistency of Myers' action for D-instantons and D-particles with the $p$-form gauge invariances, taking into account both the Dirac-Born-Infeld and the Chern-Simons parts of the action.
 
Another point of view on the gauge invariances of the action $S(X)$, which is most natural in the emerging space framework, is to start with no prejudice and study in which cases two different sets of coefficients 
$(\cn)_{n\geq 0}$ and $(c'_{i_{1}\cdots i_{n}})_{n\geq 0}$ describe the same physics and thus should be considered to be equivalent. The most general transformation laws one can consider are associated with the most general 
redefinitions of the matrix variables $X^{i}$ that are consistent with the single-trace property of the action. The set of all these transformations defines a large group $\GD$ of gauge transformations, which we may call the gauge group of D-geometry, and which we study in Section 3. This group contains the ordinary diffeomorphisms and supergravity gauge invariances but also more general, background-dependent gauge transformations that are crucial for a proper interpretation of the non-abelian D-brane actions. These general transformations will be derived and discussed in Sections 4 and 5. An example of application is to show, in Section 4, that $\GD$ acts on the $\text{AdS}_{5}\times\text{S}^{5}$ metric and five-form field strength $F_{5}$ derived in \cite{fer1} in a very simple way: the full action of $\GD$ corresponds in this case to the usual tensor transformation laws of the metric and $F_{5}$ under the group of ordinary diffeomorphisms induced by the action of $\GD$ on the set of commuting matrices. This result nicely shows that the tensor properties of 
the metric and $F_{5}$ can also be considered to be emergent properties following from the microscopic description given in \cite{fer1}. Similar results can be straightforwardly derived for the emerging geometries found in \cite{fmr,fm,fr}.

The plan of the paper is as follows. In Section 2, we study the Taylor expansion of single-trace matrix functions. In Section 3, we define and study the gauge group $\GD$. In Section 4, we derive the $\GD$ gauge transformations and apply the results to the case of D-instantons.
In Section 5, we briefly present the generalization of the discussion of Sections 2, 3 and 4 to the case of quantum mechanical actions and D-particles. Finally, we have included an Appendix on tensor symmetries in which we review calculational techniques that are heavily used throughout the main text and that we have implemented in Mathematica.
\section{The Taylor expansion}
\label{Taylorsec}

Let us start with a

\noindent\textsc{Definition:} \emph{A function $S(X)$ of $k\times k$ Hermitian matrices $X^{1},\ldots,X^{d}$ is said to be \emph{single-trace} if its expansion around an arbitrary diagonal configuration}
\be\label{Xexpdef} X^{i}=x^{i}\mathbb I + \e^{i}\ee
\emph{takes the form}
\be\label{Sexpdef} S\bigl(x\mathbb I + \e\bigr) =
\sum_{n\geq 0}\frac{1}{n!}\cn (x)\tr \e^{i_{1}}\cdots\e^{i_{n}}\, ,\ee
\emph{for a set of cyclic coefficients}
\be\label{cncyc} \cn(x)=c_{i_{n}i_{1}\cdots i_{n-1}}(x)\, .\ee
A similar definition can be given for single-trace actions $S(X)$ depending of matrix-valued fields $X^{1},\ldots X^{d}$.

\noindent Note that the cyclicity conditions on the coefficients can always be imposed, without loss of generality, from the cyclicity property of the trace. All tree-level open string effective actions are single-trace, because they can be computed from disk diagrams for which the contraction of Chan-Paton factors automatically yields a single-trace structure. 

In this Section, we are going to discuss the most general consistent form of the Taylor expansion \eqref{Sexpdef}, for an arbitrary single-trace function (or potential) $S$. The generalization to higher dimensional action is straightforward, see e.g.\ the discussion in Section \ref{qmsec}. The fact that the coefficients $\cn$ cannot be chosen arbitrarily is already clear in the trivial abelian case $k=1$, for which \eqref{Sexpdef} is the usual Taylor expansion of a function of $d$ commuting real variables. All the higher order coefficients $\cn$, $n\geq 1$, are then fixed in terms of the zeroth order coefficient as $\cn=\partial_{i_{1}\cdots i_{n}}c$.

\subsection{\label{ccondsec} The consistency conditions}

The fundamental consistency condition on the expansion \eqref{Sexpdef} is the 
invariance of the action under the simultaneous shifts
\be\label{ccond} \delta x^{i}=a^{i}\, ,\quad \delta\e^{i}=-a^{i}\mathbb I
\ee
that leave $X^{i}=x^{i} \mathbb I+\e^{i}$ unchanged. This condition ensures, at least formally, that the expansions \eqref{Sexpdef} around arbitrary points $x$ all define the same function $S$, independently of the points $x$ around which we expand.

Let us emphasize that the symmetry under the shifts \eqref{ccond} is a completely general consistency requirement and does not assume the existence of additional structures, like a metric, or even a notion of diffeomorphism invariance, on the manifold spanned by the coordinates $x$. A very similar notion of base-point independence was used in \cite{gc}, assuming the existence of a metric and diffeomorphism invariance, in order to ensure the consistency of an expansion in Riemann normal coordinates around arbitrary points.

The invariance of the action under arbitrary finite shifts \eqref{ccond} follows from its invariance under infinitesimal shifts. Taking into account \eqref{cncyc}, the invariance under infinitesimal shifts is equivalent to the conditions
\begin{align}\label{coefcond0}
&\partial_{i}c = c_{i}\, ,\\\label{coefcond}
&\partial_{i}c_{i_{1}\cdots i_{n}}=\frac{1}{n}\bigl(
c_{ii_{1}\cdots i_{n}}+ c_{ii_{2}\cdots i_{n}i_{1}}+\cdots
+c_{ii_{n}i_{1}\cdots i_{n-1}}\bigr)\quad \text{for }n\geq 1\, .\end{align}
It is convenient to rewrite these equations in the language explained in the Appendix. Let us introduce the idempotent, belonging to the group algebra of $\Sno$, defined by
\be\label{jZndef}
j_{\Zn'} = \frac{1}{n}\sum_{\sigma\in\Zn'}\sigma\, ,
\ee
where $\Zn'$ is the cyclic subgroup of $\Sno$ generated by the cycle $(2\cdots n+1)$. This idempotent acts on tensors\footnote{The fact that the coefficients $\cn$ can be considered to be tensors with respect to 
$\text{GL}(d)$ transformations will be a trivial consequence of the discussion in Section \ref{gtsec}.} of rank $n+1$ and the conditions \eqref{coefcond} are simply
\be\label{coefcond2} \partial_{i}c_{i_{1}\cdots i_{n}}=\bigl(j_{\Zn'}\cdot c\bigr)_{ii_{1}\cdots i_{n}}\, .\ee

\subsection{Simple consequences}

Let us symmetrize \eqref{coefcond} with respect to the indices $i_{1},\ldots,i_{n}$. This yields
\be\label{symc1} \partial_{i}c_{(i_{1}\cdots i_{n})}= c_{i(i_{1}\cdots i_{n})}\, .\ee
This equation can be further simplified because
\be\label{cycprop1}
c_{i(i_{1}\cdots i_{n})}=c_{(ii_{1}\cdots i_{n})}\ee
for a tensor satisfying \eqref{cncyc}.
Equation \eqref{symc1} thus implies
\be\label{symc2}\partial_{i_{1}}c_{(i_{2}\cdots i_{n+1})}=c_{(i_{1}\cdots
i_{n+1})}\, ,\ee
which can be easily solved recursively to yield
\be\label{symcoeff} c_{(i_{1}\cdots i_{n})}=\partial_{i_{1}\cdots i_{n}}c
=\frac{1}{k}\partial_{i_{1}\cdots i_{n}}S\, .\ee
This simple result is the non-abelian version of the usual Taylor expansion for functions of commuting variables.

Let us now assume that $n$ is odd and consider the completely antisymmetric combination of \eqref{coefcond} in the $n+1$ indices $i,i_{1},\ldots,i_{n}$. Since the cyclic permutations of $i_{1},\ldots,i_{n}$ are even for odd $n$, we get
\be\label{anticonda} \partial_{[i}c_{i_{1}\cdots i_{n}]}=c_{[ii_{1}\cdots i_{n}]}\, .\ee
We can now use the cyclicity of $c_{ii_{1}\cdots i_{n}}$ in its $n+1$ indices and the fact that $n+1$ is even to conclude that the right-hand side of \eqref{anticonda} vanishes and thus
\be\label{antisymcoeff} \partial_{[i_{1}}c_{i_{2}\cdots i_{n+1}]}
=0\, .\ee
These constraints generalize the conditions \eqref{anti3and5} indicated in the Introduction. In the differential form notation,
\be\label{Cndef} F^{(n)}=\frac{1}{n!}c_{[i_{1}\cdots i_{n}]}\d x^{i_{1}}\wedge\cdots\wedge\d x^{i_{n}}\, ,\ee
they are equivalent to
\be\label{dC} \d F^{(n)}=0\, .\ee
That the coefficients should satisfy such differential constraints is an interesting and quite unexpected feature of the general non-abelian Taylor expansion \eqref{Sexpdef}.

\subsection{General analysis}

To work out the most general consequences of \eqref{coefcond2}, it is convenient to decompose the cyclic tensors $\cn$ into irreducible components (see the Appendix for a general discussion of this method). Since the cyclic tensors $\cn$ are characterized by the constraint
\be\label{cycidem} \bigl(\jZn\cdot c\bigr)_{\iin}=\cn\, ,\ee
where
\be\label{jzndef} \jZn = \frac{1}{n}\sum_{\sigma\in\Zn}\sigma\ee
is the idempotent associated with
the cyclic subgroup $\Zn\subset\Sn$ generated by the cycle $(12\cdots n)$, this amounts to decomposing $\jZn$ as a sum of primitive idempotents,
\be\label{jzndec} \jZn = \sum_{a}j_{a}\, .\ee
One can then write
\be\label{cndec} \cn = \sum_{a}c(j_{a})_{i_{1}\cdots i_{n}}\, ,\ee
where the irreducible pieces are given by
\be\label{cofjadef} c(j_{a})_{i_{1}\cdots i_{n}} = 
\bigl(j_{a}\cdot c\bigr)_{\iin}\, .\ee
Solving the constraints \eqref{coefcond2} is equivalent to deriving their consequences on the irreducible tensors $c(j_{a})_{\iin}$. 

For example, the decomposition \eqref{cndec} always includes an irreducible completely symmetric tensor and, for odd $n$, an irreducible completely antisymmetric tensor. It is not difficult to show that 
\eqref{symcoeff} and \eqref{antisymcoeff} yield the most general constraints on these tensors that one can derive from \eqref{coefcond2} (that there cannot be any other constraint can be deduced from an ansatz like the one presented in \ref{ansatzssec}). 

More generally, one can apply suitable primitive idempotents to \eqref{coefcond2} to isolate irreducible terms in the decomposition of $c_{i_{1}\cdots i_{n+1}}$ and express them in terms of the derivatives $\partial_{j_{1}}c_{j_{2}\cdots j_{n+1}}$ of the lowest order coefficients.
This is what we have done to derive \eqref{symc2} or \eqref{symcoeff}.
However, it is important to realize that some irreducible pieces in $c_{i_{1}\cdots i_{n+1}}$ do not appear on the right-hand side of \eqref{coefcond2}, because they are projected out by $j_{\Zn'}$. 
For this reason, the non-abelian Taylor expansion can contain new independent coefficients at various orders, which are not expressed in terms of the lower order coefficients.

Moreover, one can note that the left-hand side of \eqref{coefcond2} is only constrained by the cyclicity in the indices $i_{1},\ldots,i_{n}$, or in other words belongs to the image of $j_{\Zn'}$, whereas the right-hand side of \eqref{coefcond2} belongs to the a priori smaller image of 
$j_{\Zn'}j_{\mathbb Z_{n+1}}$. In other words, the right-hand side of \eqref{coefcond} contains a priori less irreducible pieces than the left-hand side. This implies the vanishing of the extra irreducible pieces of the left-hand side, which yields non-trivial differential constraints on the coefficients. The most salient examples of these constraints are the conditions \eqref{antisymcoeff}. The other differential constraints that we have obtained in this way turn out to be consequences of the conditions discussed in the previous paragraph.

\subsection{\label{orderbyorderTaylor}The solution order by order}

Let us now present the general solution of \eqref{coefcond2} along the lines explained in the previous subsection. We shall expand
\be\label{Sndef} S = \sum_{n\geq 0}S_{n}\, ,\ee
where $S_{n}$ is the action at order $n$, and work order by order up to $n=5$. The calculations are tractable thank's to the 
techniques explained in the Appendix and their implementation in Mathematica.

\vfill\eject

\smallskip

\noindent\emph{First and second order}

\noindent At first order we obtain \eqref{coefcond0}. At second order, we get
\be\label{order2} \partial_{i}c_{j}=c_{ij}\, .\ee
The right-hand side is symmetric and thus we get the first example of a differential constraint, $\partial_{[i}c_{j]}=0$. This is of course not a new constraint but a consequence of the first order constraint \eqref{coefcond0}. 

\smallskip

\noindent\emph{Third order}

\noindent
At third order, the decomposition of the cyclic coefficient $c_{ijk}$ into irreducible tensors only involves the completely symmetric and completely antisymmetric pieces,
\be\label{irred3} c_{ijk}=c_{(ijk)}+c_{[ijk]}\, .\ee
Equation \eqref{coefcond2} is equivalent to 
\be\label{condorder3} \partial_{i}c_{jk}=c_{(ijk)}\, .\ee
This fixes $c_{(ijk)}$ as in \eqref{symcoeff} but leaves $c_{[ijk]}$ totally unconstrained. The coefficient $c_{[ijk]}$ is the first example of a new, independent coefficient in the Taylor expansion, with no commutative analogue. We also find differential constraints $\partial_{[i}c_{j]k}=0$ which are trivially satisfied from the lower order constraints. 

\smallskip

\noindent\emph{Fourth order}

\noindent At fourth order, the decomposition \eqref{jzndec} contains three primitive idempotents,
\be\label{irred4} j_{\mathbb Z_{4}} = j_{4}+j_{2,1,1}+j_{2,2}\, .\ee
We have labeled the primitive idempotents according to the shape of the Young tableau of the associated irreducible representation of $\text{S}_{4}$, an idempotent $j_{p_{1},p_{2},\ldots}$ corresponding to a Young tableau having $p_{i}$ boxes in the $i^{\text{th}}$ row. For example 
$j_{4}$, $j_{2,1,1}$ and $j_{2,2}$ are associated with 
$\Yboxdim6pt\yng(4)$, $\Yboxdim 6pt\yng(2,1,1)$ and $\Yboxdim6pt\yng(2,2)$ respectively. Acting on the cyclic coefficients with these idempotents, we get the explicit formulas for the irreducible tensors,
\begin{align}\label{tensorj4} c(j_{4})_{ijkl}& = c_{(ijkl)}\, ,\\
\label{tensorj211}
c(j_{2,1,1})_{ijkl} &= \frac{3}{2}\bigl( c_{i[jkl]} + c_{k[lij]}\bigr)
=\frac{1}{2}\bigl(c_{ijkl}-c_{ilkj}\bigr)
\, ,\\\label{tensorj22}
c(j_{2,2})_{ijkl} &= \frac{2}{3}\bigl( c_{[ij][kl]} + c_{[il][kj]}\bigr)\, ,\end{align}
or, conversely,
\be\label{tensor4inv} c_{(ijkl)}=c(j_{4})_{ijkl}\, ,\ c_{i[jkl]}=
c(j_{2,1,1})_{i[jkl]}\, ,\ c_{[ij][kl]}= c(j_{2,2})_{[ij][kl]}\, .\ee
It is interesting to note that the coefficient $c_{i[jkl]}$ can be fully characterized by its antisymmetry in the last three indices and the additional condition $c_{[ijkl]}=0$, whereas the symmetries of 
$c_{[ij][kl]}$ precisely match the symmetries of the Riemann curvature tensor, including 
\be\label{riemsym} c_{[ij][kl]}=c_{[kl][ij]}\, ,\quad
c_{[ij][kl]}+c_{[ik][lj]}+c_{[il][jk]}=0\, .\ee
The fourth order action thus reads
\begin{align}\nonumber S_{4}& = \frac{1}{4!}\bigl(c(j_{4})_{ijkl} + 
c(j_{2,1,1})_{ijkl} + c(j_{2,2})_{ijkl}\bigr)\tr\e^{i}\e^{j}\e^{k}\e^{l}
\\\label{s4gen}
& = \frac{1}{4!}c_{(ijkl)}\tr\e^{(i}\e^{j}\e^{k}\e^{l)}+ \frac{1}{8}c_{i[jkl]}\tr\e^{i}\e^{[j}\e^{k}\e^{l]}+\frac{1}{72}c_{[ij][kl]}\tr[\e^{i},\e^{j}][\e^{k},\e^{l}]\, .
\end{align}

The consequences of equation \eqref{coefcond2} for $n=3$ can then be straightforwardly studied. The coefficient $c_{(ijkl)}$ is fixed as in \eqref{symcoeff} whereas $c_{i[jkl]}$ is expressed in terms of the third order $c(j_{1,1,1})$ as
\be\label{solcond4} c_{i[jkl]}= \partial_{i}c_{[jkl]}\, .
\ee
On the other hand, the coefficient $c_{[ij][kl]}$ is left unconstrained and thus represent a new independent irreducible tensor parametrizing the Taylor expansion. Finally, \eqref{solcond4} implies the differential constraint 
\eqref{antisymcoeff} at $n=3$, since $c_{[ijkl]}$ automatically vanishes due to the cyclicity condition. 

\smallskip

\noindent\emph{Fifth order}

\noindent At fifth order, the cyclic coefficient has six irreducible pieces, according to the decomposition
\be\label{irred5} j_{\mathbb Z_{5}}= j_{5}+ j_{2,2,1}+j_{3,2}
+j_{3,1,1}+ j'_{3,1,1} + j_{1,1,1,1,1}\ee
in which the Young tableau $\Yboxdim6pt\yng(3,1,1)$ occurs twice. Acting on the cyclic coefficients with the primitive idempotents appearing in \eqref{irred5}, we obtain the following explicit expressions for the irreducible tensors,
\begin{align}\label{tensorj5} &c(j_{5})_{ijklm}=c_{(ijklm)}\, ,\\
\label{tensorj221} &c(j_{2,2,1})_{ijklm} = \frac{1}{4}\bigl( c_{ijklm}
+ c_{ikmjl} + c_{iljmk} + c_{imlkj}\bigr) - c_{(ijklm)}\, ,\\
\label{tensorj32} &c(j_{3,2})_{ijklm} = \frac{1}{4}\bigl( c_{ijklm}
- c_{ikmjl} - c_{iljmk} + c_{imlkj}\bigr) - c_{[ijklm]}\, ,\\
\label{tensorj311} &c(j_{3,1,1})_{ijklm} = 
\frac{1}{5}\bigl(c_{[ij][kl]m}-c_{[kl][ij]m} + \text{circular\ perm.}\bigr)
\, ,\\\label{tensorj311p} &c(j'_{3,1,1})_{ijklm}  =
\frac{1}{2}\bigl( c_{ijklm} - c_{imlkj}\bigr)
-c(j_{3,1,1})_{ijklm}\, ,\\\label{tensorj11111}&
c(j_{1,1,1,1,1})_{ijklm}  = c_{[ijklm]}\, .
\end{align}
The right-hand side of \eqref{tensorj311} contains eight additional terms obtained by circular permutations of the indices $ijklm$ of the two terms that we have explicitly written down.

It turns out that equation \eqref{coefcond2} for $n=4$ fixes $c(j_{5})$ as in \eqref{symcoeff} as well as $c(j_{2,2,1})$, $c(j_{3,2})$ and $c(j'_{3,1,1})$ in terms of the lower order coefficients. Explicitly, one finds
\begin{align}\label{j221plusj32sol} & c(j_{2,2,1}+j_{3,2})_{ijklm}
= \frac{4}{3}\bigl( \partial_{i}c_{[jk][lm]} + \partial_{j}c_{[kl][mi]}
+\partial_{k}c_{[lm][ij]} + \partial_{l}c_{[mi][jk]} + \partial_{m}
c_{[ij][kl]}\bigr)\, ,\\\label{j221minusj32sol}
& c(j_{2,2,1}-j_{3,2})_{ijklm}
= \frac{4}{3}\bigl( \partial_{i}c_{[jl][km]} + \partial_{j}c_{[li][km]}
+\partial_{k}c_{[il][jm]} + \partial_{l}c_{[ki][jm]} + \partial_{m}
c_{[ik][jl]}\bigr)\, ,\\
\label{j311psol}& c(j'_{3,1,1})_{ijklm} =
\frac{3}{5}\bigl(3\,\partial_{ij}c_{[klm]}+ 3\,\partial_{il}c_{[jkm]}
+ 3\,\partial_{kl}c_{[ijm]} + \partial_{ik}c_{[jlm]}+\partial_{jk}c_{[iml]}
+\partial_{jl}c_{[ikm]}\bigr)\, . 
\end{align}
The tensors $c(j_{3,1,1})$ and $c(j_{1,1,1,1,1})$ are the new tensors appearing at order five; the consistency conditions \eqref{coefcond2} do not relate them to lower order coefficients. Moreover, one can check explicitly that the differential constraints on the fifth order coefficients implied by \eqref{coefcond2} for $n=5$ are all consequences of \eqref{symcoeff}, \eqref{j221plusj32sol} and \eqref{j311psol}, except for the condition \eqref{antisymcoeff} on $c_{[ijklm]}$.

\smallskip

\noindent\emph{Sixth and higher orders}

\noindent At sixth order, we get twenty irreducible tensors, fourteen of which are fixed by \eqref{coefcond2} and six are new. And so on and so forth at higher and higher orders. It is possible to obtain explicit formulas using the computer, but they are complicated and not particularly useful. In particular, for most string theory applications, the knowledge of the action up to order five is sufficient, since it allows to derive the full supergravity background \cite{fer1,fmr}, see Section \ref{Myerssec}. 

We now turn to a slightly less rigorous but possibly more illuminating discussion, based on a general ansatz for the solution of \eqref{coefcond2}.

\subsection{\label{ansatzssec} A convenient ansatz for the solution}

It is actually very easy to write solutions to the conditions \eqref{coefcond2}, based on the following two very simple remarks:

\noindent (i) Commutators $[\e^{i},\e^{j}]=[X^{i},X^{j}]$, commutators of commutators $[\e^{i},[\e^{j},\e^{k}]]$, etc, are automatically invariant under the transformation \eqref{ccond}.

\noindent (ii) For any ordinary function $f(x^{1},\ldots,x^{d})$ of commuting variables, the series
\be\label{fhatdef} \hat f = \sum_{n\geq 0}\frac{1}{n!}\partial_{i_{1}\cdots i_{n}}f(x)\,\e^{i_{1}}\cdots\e^{i_{n}}\ee
is automatically invariant under the transformation \eqref{ccond}. In particular, we can consider that it defines a function $\hat f(X^{1},\ldots,X^{d})$ of the matrices $X^{1},\ldots,X^{d}$, with value in the space of Hermitian matrices.

The most general solution of \eqref{coefcond2}, up to order five, can then be reproduced by the following simple ansatz, which manifestly satisfies the consistency conditions of Section \ref{ccondsec},
\be\label{ansatz} S(X) = \tr\Bigl\{\hat s(X) + \hat s_{ij}(X)
\bigl[\e^{i},\e^{j}\bigr] + \hat s_{ijkl}(X)\bigl[\e^{i},\e^{j}\bigr]
\bigl[\e^{k},\e^{l}\bigr] +
\hat s_{ijklm}(X)\bigl[\e^{i},\e^{j}\bigr]\bigl[\e^{k},[\e^{l},\e^{m}]
\bigr]\Bigr\} .\ee
The symmetry properties of the commutators in \eqref{ansatz} allow us to constrain the coefficients $s_{ij}$, $s_{ijkl}$ and $s_{ijklm}$ as
\begin{align}\label{coefshatcons1} & s_{ij}=-s_{ij}\, ,\\
\label{coefshatcons2} & s_{ijkl}=-s_{jikl}=-s_{ijlk}\, ,\\
\label{coefshatcons3} & s_{ijklm}=-s_{jiklm}=-s_{ijkml}\, ,\ 
s_{ijklm}+s_{ijlmk}+ s_{ijmkl} = 0\, .
\end{align}
By construction, the expansion of \eqref{ansatz} in powers of $\e$, using \eqref{fhatdef}, yields coefficients $\cn$ that automatically satisfy \eqref{coefcond2} and thus all the relations discussed in the previous subsection.

Up to order two, only $\hat s$ contributes, with $c=s$, $c_{i}=\partial_{i} s$, $c_{ij}=\partial_{ij}s$. At order three, on top of $\hat s$ which yields the completely symmetric coefficient $c_{(ijk)}=\partial_{ijk}s$, the term in $\hat s_{ij}$ yields 
\be\label{cantipot} c_{[ijk]}=12\,\partial_{[i}s_{jk]}\, .\ee
This formula automatically implements the differential constraint \eqref{antisymcoeff}, the three-form being expressed as
$F^{(3)}=4\d s^{(2)}$ if $s^{(2)}$ is the two-form with components 
$s_{[ij]}$. The fourth order action derived from \eqref{ansatz} reads
\be\label{ans4} S_{4}= \frac{1}{4!}\partial_{ijkl}s\tr\e^{i}\e^{j}\e^{k}\e^{l} + \frac{1}{2}\partial_{ij}s_{kl}\tr\e^{i}\e^{j}[\e^{k},\e^{l}] + s_{ijkl}\tr [\e^{i},\e^{j}][\e^{k},\e^{l}]\, .\ee
Using the cyclicity of the trace, it is easy to check that the tensor $\partial_{ij}s_{kl}$ actually enters only via the components $\partial_{i[j}s_{kl]}$, consistently with \eqref{cantipot}, \eqref{solcond4} and \eqref{tensorj211}.
Taking into account \eqref{coefshatcons2}, the tensor $s_{ijkl}$ a priori contains three new irreducible components, but at order four only the cyclic combination appears, which yields the unique irreducible piece
\be\label{j22s4rel} c(j_{2,2})_{ijkl} = 24\bigl( s_{ijkl}+ s_{jkli}
+ s_{klij} + s_{lijk}\bigr)\, ,\ee
consistently with the discussion in \ref{orderbyorderTaylor}.
The fifth order action reads
\begin{multline}\label{ans5} S_{5}=\frac{1}{5!}\partial_{ijklm}s
\tr\e^{i}\e^{j}\e^{k}\e^{l}\e^{m} + \frac{1}{3!}\partial_{ijk}s_{lm}
\tr\e^{i}\e^{j}\e^{k}[\e^{l},\e^{m}]\\
+ \partial_{i}s_{jklm}\tr\e^{i}[\e^{j},\e^{k}][\e^{l},\e^{m}] + 
s_{ijklm}\tr [\e^{i},\e^{j}][\e^{k},[\e^{l},\e^{m}]] \, .
\end{multline}
One can check that only the components of the form
$\partial_{ij[k}s_{lm]}$ of $\partial_{ijk}s_{lm}$ contribute, consistently with \eqref{j311psol} and \eqref{cantipot}. The term in $\partial_{i}s_{jklm}$ yields $c(j_{2,2,1}\pm j_{3,2})$, consistently with \eqref{j221plusj32sol}, \eqref{j221minusj32sol}, \eqref{j22s4rel} and the last equation in \eqref{tensor4inv}. It also yields the new irreducible tensor
\be\label{j11111s4rel} c(j_{1,1,1,1,1})_{ijklm}= c_{[ijklm]}= 480\,
\partial_{[i}s_{jklm]}\, ,\ee
in a form that manifestly satisfies the constraint \eqref{antisymcoeff}.
Finally, the new tensor $c(j_{3,1,1})$ picks contributions from $\partial_{ijk}s_{lm}$, $\partial_{i}s_{jklm}$ and $s_{ijklm}$,
\begin{multline}\label{j311s245rel} c(j_{3,1,1})_{ijklm} = \frac{12}{5}\bigl(
\partial_{ij[k}s_{lm]} + \partial_{il[j}s_{km]} +
\partial_{kl[i}s_{jm]} - 3\,\partial_{ik[j}s_{lm]}
+ 3\, \partial_{jk[i}s_{lm]} - 3\,\partial_{jl[i}s_{km]}\bigr)\\
+ 48\bigl(\partial_{i}s_{jklm}-\partial_{i}s_{lmjk} + \text{circ.\ perm.\ on }(ijklm)\bigr)\\
+ 96\bigl(s_{ijklm}-s_{ijmkl} + \text{circ.\ perm.\ on }(ijklm)\bigr)\, .
\end{multline}
\subsection{Reality condition}

In some cases, for example if $S(X)$ is the effective potential in a Minkowskian world\-volume action, it may be natural to impose a reality condition
\be\label{realS} S(X)^{*} = S(X)\, ,\ee
which is equivalent to
\be\label{realcyclic} c_{\iin}^{*}=c_{i_{n}\cdots i_{1}}\ee
on the coefficients. If we denote by $\sigma$ the permutation such that $\sigma(k) = n-k+1$, it is not difficult to check that $\sigma\jZn =
\jZn\sigma$. The group algrebra elements
\be\label{jznpmdef} \jZn^{\pm} = \frac{1\pm\sigma}{2}\jZn
\ee
are then Hermitian idempotents corresponding to orthogonal projectors on orthogonal subspaces. The decomposition \eqref{jzndec} can be written as
\be\label{decreality} \jZn = \jZn^{+} + \jZn^{-} = 
\sum_{a}j_{a}^{+} + \sum_{b}j_{b}^{-}\, ,\ee
with $\sigma j_{a}^{\pm} = \pm j_{a}^{\pm}$. The associated irreducible tensors in \eqref{cndec} can thus be chosen such that
\be\label{irrrealcond} c(j_{a}^{\pm})_{i_{n}\cdots i_{1}} = \pm
c(j_{a}^{\pm})_{\iin}\, .\ee
The reality condition \eqref{realcyclic} is then equivalent to imposing that the irreducible tensors $c(j_{a}^{+})$ and $c(j_{a}^{-})$ are real and purely imaginary respectively. Let us note that the reality constraints obtained in this way are automatically consistent with the consistency conditions \eqref{coefcond2}, since the invariance of $S(X)$ under the shifts \eqref{ccond} implies the invariance of $S(X)^{*}$ under the same shifts.
In the order by order analysis performed in Section \ref{orderbyorderTaylor}, one finds that the irreducible tensors are all real up to order two, together with $c(j_{3})$, $c(j_{4})$, $c(j_{2,2})$, $c(j_{5})$, $c(j_{2,2,1})$, $c(j_{3,2})$ and $c(j_{1,1,1,1,1})$, whereas $c(j_{1,1,1})$, $c(j_{2,1,1})$, $c(j_{3,1,1})$ and $c(j'_{3,1,1})$ are purely imaginary. 

\subsection{Summary of results}

Through its Taylor expansion \eqref{Sexp}, a single-trace function of matrices $S(X)$ can be characterized by an infinite set of irreducible tensors 
\be\label{Sequiv} S(X)\equiv\bigl(c(x), c_{[ijk]}(x), c_{[ij][kl]}(x), 
c(j_{3,1,1})_{ijklm}(x),c_{[ijklm]}(x), \ldots\bigr)\, .\ee
These tensors correspond to the irreducible pieces of the coefficients in the expansion that are not expressed as derivatives of lower order coefficients by the consistency conditions \eqref{coefcond2}. Moreover, the completely antisymmetric tensors of odd order appearing in \eqref{Sequiv} must satisfy \eqref{antisymcoeff}, i.e.\ they are associated with closed differential forms. If the function $S(X)$ is real, the irreducible tensors must be real or purely imaginary according to their parity in \eqref{irrrealcond}.

\subsection{The example of Myers D-instanton action}\label{Myerssec}

A very natural example of a single-trace matrix action like our $S(X)$ is the effective action for D-instantons in type IIB string theory. Each Hermitian matrix $X^{i}$ is associated in this case with a Euclidean spacetime dimension and thus $d=10$. The size $k$ of the matrices is identified with the number of D-intantons. The single-trace structure of the action is a consequence of the small string coupling approximation, in which the action can be computed from open string diagrams having only one boundary. In this context, the tensors in \eqref{Sequiv} are naturally identified with the non-trivial closed string background fields in which the D-instantons can move. 

Myers \cite{Myers} has proposed a general formula for $S(X)$, using in particular constraints from T-duality.
Myers' action is the sum of Dirac-Born-Infeld and Chern-Simons terms,
\be\label{Mac1} S(X) = S_{\text{DBI}}(X) + S_{\text{CS}}(X)\, .\ee
The Dirac-Born-Infeld part is given by
\be\label{SBI} S_{\text{DBI}} = 2\pi\Str e^{-\phi}\sqrt{\det 
\bigl(\delta^{i}_{j} + i\ls^{-2}[\epsilon^{i},\epsilon^{k}]
(G_{kj} + B_{kj})\bigr)}\, ,\ee
where $\phi$, $G_{ij}$ and $B_{ij}$ are the usual dilaton, metric and Kalb-Ramond two-form of the Neveu-Schwarz sector and $\ls$ is the string length. The fields are evaluated at $X=x+\e$. The determinant acts on the indices $i,j$ (not on the $\text{U}(k)$ indices of the matrices $\epsilon$). The Str is an appropriate symmetrized trace on the $\text{U}(k)$ indices whose precise definition is given in \cite{Myers} (and which should provide the correct ordering for the action up to order five in the expansion \eqref{Sexp}, but not beyond). The Chern-Simons part of the action is given by
\be\label{SCS} S_{\text{CS}} = 2i\pi\Str e^{i\ls^{-2}i_{\epsilon}i_{\epsilon}}\sum_{q\geq 0} C_{2q}\wedge e^{B}\vert_{0-\text{form}}\, ,\ee
where we keep only the 0-form part in the right-hand side, the $C_{2q}$ are the Ramond-Ramond forms and $i_{\epsilon}$ the inner product. It is straightforward to check that, up to order five, \eqref{SBI} and \eqref{SCS} yield an action of the form \eqref{ansatz}, with
\begin{align}\label{smyers} s & = -2i\pi\bigl( ie^{-\phi} - C_{0}\bigr)
= -2i\pi\tau\, ,\\\label{sijmyers} s_{ij}&=-\frac{\pi}{\ls^{2}}
\bigl(\tau B - C_{2}\bigr)_{ij}\, ,\\\label{sijklmyers}
\begin{split}
s_{ijkl} & = \frac{\pi}{4\ls^{4}}e^{-\phi}\bigl(G_{jk}G_{il}-G_{ik}G_{jl}
-B_{jk}B_{il}+B_{ik}B_{jl}-B_{ij}B_{kl}\bigr)\\
&\hskip 5cm
-\frac{i\pi}{4\ls^{4}}\bigl(C_{4}+C_{2}\wedge B + \frac{1}{2}C_{0}
B\wedge B\bigr)_{ijkl}\, ,
\end{split}
\\\label{sijklmmyers}
s_{ijklm} & = 0\, .
\end{align}
The independent irreducible tensors entering into the expansion of $S$ follow. We immediately get
\be\label{c0c3myers} c = -2i\pi\tau\, ,\quad c_{[ijk]}=
-\frac{12\pi}{\ls^{2}}\partial_{[i}( \tau B - C_{2})_{jk]}\, .\ee
At order four, only the cyclic combination of the $s_{ijkl}$ enters, which eliminates the $BB$ and antisymmetric terms in \eqref{sijklmyers}, yielding
\be\label{c4myers} c(j_{2,2})_{ijkl}=-\frac{12\pi}{\ls^{4}}e^{-\phi}
\bigl(2G_{ik}G_{jl}-G_{ij}G_{kl}-G_{il}G_{jk}\bigr)\, ,\ee
or equivalently from \eqref{tensor4inv}
\be\label{c4myersbis} c_{[ij][kl]}=-\frac{18\pi}{\ls^{4}}e^{-\phi}
\bigl(G_{ik}G_{jl}-G_{il}G_{jk}\bigr)\, .\ee
At order five, the vanishing of $s_{ijklm}$ implies that there is only one new independent irreducible tensor at this order (instead of two for a generic matrix action), given by \eqref{j11111s4rel} as
\be\label{c5myers} c_{[ijklm]} =
-\frac{120 i \pi}{\ls^{4}}
\partial_{[i}\bigl( C_{4} + C_{2}\wedge B - \frac{1}{2}\tau B\wedge B
\bigr)_{jklm]}\, .\ee
The would-be new independent tensor $c(j_{3,1,1})$ is expressed in the present case in terms of lower order coefficients according to \eqref{j311s245rel} for $s_{ijklm}=0$. Let us note that \eqref{c0c3myers}, \eqref{c4myers} and \eqref{c5myers} show that the full set of supergravity fields is encoded into the irreducible tensors appearing in the expansion \eqref{Sexp} up to order five.

The formulas \eqref{c0c3myers}, \eqref{c4myersbis} and \eqref{c5myers}, derived from the Myers' action, correspond to the first terms in an infinite derivative expansion in powers of $\ls^{2}$. This derivative expansion can in principle be obtained by computing open string disk diagrams. Terms of order $n$ are generated when $n$ open string vertex operators are inserted on the boundary of the disk, together with closed string vertex operators in the bulk. These calculations are of course extremely difficult, in particular at high orders, and only a few examples can be found in the literature, e.g.\ in \cite{couplings}.

However, the consistency conditions that we have studied above are exact and thus constrain the form of the action to all orders in $\ls^{2}$. For example, from Myers', we find that the differential forms $F^{(1)}$, $F^{(3)}$ and $F^{(5)}$ defined in \eqref{Cndef} are given by
\be\label{F1F3F5myers} F^{(1)} = -2i\pi\d\tau\, ,\ F^{(3)} = -\frac{4\pi}{\ls^{2}}\d\bigl(\tau B - C_{2}
\bigr)\, ,
\ F^{(5)} = -\frac{24i\pi}{\ls^{4}}\d \bigl(C_{4}+ C_{2}\wedge B
-\frac{1}{2}\tau B\wedge B\bigr)\, .\ee
The fact that these forms are locally exact is not an accident of the leading $\ls^{2}$ expansion, but instead a consequence of the general consistency conditions \eqref{antisymcoeff}. This property will thus remain valid to all orders in $\ls^{2}$ and even at finite $\ls^{2}$. Explicit examples are worked out in \cite{fmr}. We could then use \eqref{F1F3F5myers} to actually \emph{define} what 
we mean by $\tau$, $B$, $C_{2}$ and $C_{4}$ to all orders in $\ls^{2}$. The general gauge transformations of the $p$-form potentials defined in this way will be discussed in Section \ref{gtsec}.

On the other hand, there are features of the Myers' action than will not remain valid to all orders in $\ls^{2}$, because they are not protected by the general consistency conditions. For example, the fourth order coefficient $c_{[ij][kl]}$ factorizes in terms of a second rank metric tensor in \eqref{c4myersbis}. However, we have seen that the only general constraint on this coefficient is that it should have the same symmetries as the Riemann curvature tensor, including \eqref{riemsym}. This is not enough to ensure the existence of a factorized formula like \eqref{c4myersbis}. Such a special form for $c_{[ij][kl]}$ will thus be preserved only in exceptional situations, probably only when the $\ls^{2}$ corrections vanish, which occurs for the maximally supersymmetric $\text{AdS}_{5}\times\text{S}^{5}$ background \cite{fer1}.

\section{\label{ggsec} The gauge group}

In the abelian case, $k=1$, two actions $S$ and $S'$ related by a simple reparameterization of the spacetime coordinates, i.e.\ such that $S'(x')=S(x)$ for a diffeomorphism $x\mapsto x'$, should of course be considered to be physically equivalent. In the non-abelian case, the $d$ coordinates $x^{i}$ are promoted to $k\times k$ Hermitian matrices $X^{i}$. One may then be tempted to consider the group of diffeomorphisms acting on the $d k^{2}$ real independent entries of the matrix coordinates $X^{i}$. However, this huge group of transformations is not really interesting.  
It is much more fruitful to take into account basic properties of our matrix actions, which emerge naturally from string theory.

First, there is a gauge group $\text{U}(k)$ acting on the matrices as $X\mapsto UXU^{-1}$. This gauge group is automatically present in the microscopic open string description. We wish to restrict the allowed transformations $X\mapsto X'$ to be compatible with the $\text{U}(k)$ action. Second, at small $\gs$ (which, in the microscopic gauge-theoretic description, corresponds to a large $N$ limit), the effective actions are automatically single-trace. It is thus also very natural to restrict ourselves to transformations that respect the single-trace structure. These considerations yield the following definitions.

\noindent\textsc{Definition 1}: \emph{The quantum gauge group
${\mathscr G}$ of D-geometry is the subgroup of the group of diffeomorphisms acting on the $d k^{2}$ independent real entries of the matrix coordinates $X^{i}$ such that, for any $F\in {\mathscr G}$, there exists a
$\text{U}(k)$ automorphism $\rho$ such that}
\be\label{compatible} F(UXU^{-1})^{i}= \rho(U)F(X)^{i}\rho(U)^{-1}\ee
\emph{for any $U\in\text{U}(k)$.}

\noindent The simple transformations
\be\label{innertrans} X'^{i}=U_{0}X^{i}U_{0}^{-1}\, ,\ee
for $U_{0}\in\text{U}(k)$, belong to ${\mathscr G}$, with associated (inner) automorphism $\rho(U)=U_{0}UU_{0}^{-1}$. Another simple transformation belonging to ${\mathscr G}$ is complex conjugation,
\be\label{outertrans} X'=(X^{i})^{*}\, ,\ee
with associated automorphism $\rho(U)=U^{*}$. Since complex conjugation is actually the only outer automorphism of $\text{U}(k)$, we see that ${\mathscr G}$ is generated by \eqref{innertrans}, \eqref{outertrans} and the transformations satisfying the simple constraint
\be\label{compatible2} F(UXU^{-1})^{i}= U F(X)^{i}U^{-1}\, .\ee

\noindent\textsc{Definition 2}: \emph{The classical gauge group of D-geometry $\GD$ is the subgroup of ${\mathscr G}$ preserving the single-trace structure of a matrix action, i.e., it corresponds to the transformations $X\mapsto X'$ such that, if $S$ is single-trace, then $S'$ defined by $S'(X') = S(X)$ is also single-trace.}

\noindent Let us note that \eqref{innertrans} and \eqref{outertrans} belong to $\GD$ and thus we can restrict ourselves without loss of generality to the transformations of $\GD$ satisfying the simpler condition \eqref{compatible2}. Our aim in the present Section is to provide an explicit description of $\GD$ and discuss some of its elementary structural properties.

\subsection{The consistency conditions}\label{ccongroupsec}

A transformation $\gamma\in\GD$ satisfying \eqref{compatible2} can be described by the set of coefficients
$\gn$ that appear in the expansion
\be\label{gamexp} X'^{i}=\gamma^{i}(X)=\gamma^{i}(x\mathbb I+\epsilon) = \sum_{n\geq 0}\frac{1}{n!}\gamma^{i}_{i_{1}\cdots i_{n}}(x)\e^{i_{1}}\cdots\e^{i_{n}}\, .\ee
Let us note that this is the most general form of the expansion that is compatible with both \eqref{compatible2} and the single-trace restriction. In particular, if the expansion contained explicit traces, then multi-trace terms would be produced when acting on a single-trace matrix action, which is forbidden. The hermi\-ti\-city of the matrix coordinates $X^{i}$ imply that the $\gn$ must satisfy a reality constraint
\be\label{realc} (\gn)^{*} = \gamma^{i}_{i_{n}i_{n-1}\cdots i_{1}}\, .\ee
Introducing the real and imaginary parts of the coefficients,
\be\label{abdef} \an=\re\gn\, ,\quad \bn=\im\gn\, ,\ee
this is equivalent to the conditions
\begin{align}\label{realcr} \an & = \alpha^{i}_{i_{n}i_{n-1}\cdots i_{1}}\, ,\\\label{realci}
\bn & = -\beta^{i}_{i_{n}i_{n-1}\cdots i_{1}}\, .
\end{align}

The most general consistent expansion \eqref{gamexp} for $\gamma\in\GD$ can be found by a rather straightforward generalization of the approach used in Section \ref{Taylorsec} to characterize the non-abelian Taylor expansions of single-trace functions. The fundamental consistency condition on the expansion \eqref{gamexp} is the invariance under the shifts \eqref{ccond}, which yields
\be\label{gamcond} \partial_{j}\gn = \frac{1}{n+1}\sum_{k=1}^{n+1}
\gamma^{i}_{i_{1}\cdots i_{k-1}ji_{k}\cdots i_{n}}\, .\ee
These equations are most conveniently written as
\be\label{gamcond2} \partial_{i_{1}}\gamma^{i}_{i_{2}\cdots i_{n+1}}
= \bigl(J_{n+1}\cdot\gamma^{i}\bigr)_{i_{1}\cdots i_{n+1}}\, ,\ee
in terms of the element
\be\label{Jdef} J_{n}=\frac{1}{n}\sum_{k=1}^{n}(12\cdots k)\ee
of the group algebra $\CSn$. By taking the real and imaginary parts, we get
\begin{align}
\label{acond} \partial_{i_{1}}\alpha^{i}_{i_{2}\cdots i_{n}}
& = \bigl(J_{n}\cdot\alpha^{i}\bigr)_{i_{1}\cdots i_{n}}\, ,\\
\label{bcond}
\partial_{i_{1}}\beta^{i}_{i_{2}\cdots i_{n}}
& = \bigl(J_{n}\cdot\beta^{i}\bigr)_{i_{1}\cdots i_{n}}\, .
\end{align}
Equations \eqref{acond}, \eqref{bcond} and \eqref{realcr}, \eqref{realci} 
are the analogues of the constraints \eqref{coefcond2} and \eqref{cncyc} used in Section \ref{Taylorsec}.

Since the upper index in $\alpha$, $\beta$ or $\gamma$ does not play any 
r\^ole in the consistency conditions \eqref{realcr}, \eqref{realci}, \eqref{acond} and \eqref{bcond}, we are going to suppress it in the following subsections in order to simplify the notation. 

\subsection{Simple consequences}

Let us symmetrize \eqref{acond} and \eqref{bcond} with respect to the lower indices. We get
\begin{align}\label{asym} \alpha_{(i_{1}\cdots i_{n})}& = 
\partial_{i_{1}\cdots i_{n}}\alpha\, ,\\\label{bsym}
\beta_{(i_{1}\cdots i_{n})} & = 0\, ,
\end{align}
which are the analogues of \eqref{symcoeff}. Similarly, by antisymmetrizing with respect to the lower indices, we obtain the relations
\begin{align}\label{dA}
& \d A^{(4p)} = A^{(4p+1)}\, ,\quad \d A^{(4p+1)}=0
=A^{(4p+2)}=A^{(4p+3)}\, ,\\\label{dB} &
\d B^{(4p-1)}= B^{(4p)} = B^{(4p+1)}=0\, ,\quad \d B^{(4p+2)}=
B^{(4p+3)}
\end{align}
on the forms
\be\label{AnBndef} A^{(n)}=\frac{1}{n!}\alpha_{[i_{1}\cdots i_{n}]}
\d x^{i_{1}}\wedge\cdots\wedge\d x^{i_{n}}\, ,\quad
B^{(n)}=\frac{1}{n!}\beta_{[i_{1}\cdots i_{n}]}
\d x^{i_{1}}\wedge\cdots\wedge\d x^{i_{n}}\ee
built from the totally antisymmetric coefficients. Equations \eqref{dA} and \eqref{dB} are the analogues of \eqref{dC}.

\subsection{The solution order by order}

Let us present the general solution up to order four. The derivations follow the same lines as in Section \ref{Taylorsec} and are based on the principles outlined in the Appendix. More details on an illustrative example can be found in 
Section \ref{Appsamplesec}.

\smallskip

\noindent\emph{First and second order}

\noindent For the real part coefficients, all the constraints at orders one and two are included in \eqref{asym} and the conditions $A^{(1)}=\d A^{(0)}$ and $A^{(2)}=0$ from \eqref{dA}, which yield
\be\label{o12a} \alpha_{i}=\partial_{i}\alpha\, , \quad
\alpha_{(ij)}= \partial_{ij}\alpha\, ,\quad\alpha_{[ij]} = 0\, .\ee
On the other hand, the zeroth and first order imaginary part coefficients vanish, $\beta=\beta_{i}=0$. At order two, since $\beta_{(ij)}=0$ we get one unconstrained irreducible tensor $\beta_{[ij]}$ corresponding to the form $B^{(2)}$ in \eqref{AnBndef}.

To summarize, up to order two, $\gamma\in\GD$ is parametrized by an ordinary diffeomorphism $\alpha$ and a two-form $b=B^{(2)}$, $b_{ij}=\beta_{[ij]}=\beta(j_{1,1})_{ij}$.

\smallskip

\noindent\emph{Third order}

\noindent The condition $\alpha_{ijk}=\alpha_{kji}$ implies that
$\alpha_{[ijk]}=0$ and thus we have a decomposition in irreducible tensors of the form
\be\label{deca3} \alpha_{ijk}= \alpha(j_{3})_{ijk} + \alpha(j_{2,1})_{ijk}\, .\ee
The coefficient $\alpha(j_{3})_{ijk}=\alpha_{(ijk)}$ is fixed by \eqref{asym}, whereas
\be\label{aj21} \alpha(j_{2,1})_{ijk} = \frac{1}{3}\bigl(2\alpha_{ijk}
-\alpha_{ikj}-\alpha_{jik}\bigr)=\frac{2}{3}\bigl(\alpha_{i[jk]}+\alpha_{[ij]k}\bigr)\ee
is left unconstrained. If we define
\be\label{uj21rel} u_{ijk}=\a_{i[jk]}=\a(j_{2,1})_{i[jk]}\, ,\ee
then one can show that the irreducible tensor $u$ is characterized by the following symmetry properties,
\be\label{usym} u_{ijk}=-u_{ikj}\, ,\quad
u_{ijk}+u_{jki}+u_{kij}=0\, .\ee
On the other hand, the condition $\beta_{ijk}=-\beta_{kji}$ allows two  irreducible components for the imaginary part coefficients. The completely antisymmetric irreducible piece $\beta_{[ijk]}$ is fixed by \eqref{dB},
\be\label{B3rel} \beta_{[ijk]}=3\,\partial_{[i}\beta_{jk]}\, ,\ee
and the other component, given by
\be\label{bj21}\beta(j_{2,1})_{ijk}=\frac{1}{3}\bigl( 
2\b_{ijk}+\b_{ikj}+\b_{jik}\bigr)=\frac{2}{3}\bigl(\b_{i(jk)}+\b_{(ij)k}
\bigr)\, ,\ee
is fixed by the consistency condition \eqref{bcond} at $n=3$ to be
\be\label{solbj21} \beta(j_{2,1})_{ijk} = \frac{1}{2}\bigl(
\partial_{i}\b_{[jk]}-2\partial_{j}\beta_{[ki]}+\partial_{k}\b_{[ij]}\bigr)\, .\ee
Note that we are using the same convenient notation as in Section \ref{Taylorsec} for the primitive idempotents, which are denoted according to the shape of the associated Young tableau. However, this notation is ambiguous, since two distinct idempotents can be associated with the same Young tableau. For example, the idempotent $j_{2,1}$ in \eqref{bj21} is clearly not the same as the one in \eqref{aj21}, as the formulas for $\alpha(j_{2,1})$ and $\beta(j_{2,1})$ show, but we are using the same notation because they are both associated with the same tableau $\Yboxdim 6pt\yng(2,1)$. 

\smallskip

\noindent\emph{Fourth order}

\noindent The condition $\a_{ijkl}=\a_{lkji}$ allow six irreducible pieces,
\begin{align}\label{decaj4} &\alpha(j_{4})_{ijkl}  = \alpha_{(ijkl)}\, ,
\\\label{decaj31} &\alpha(j_{3,1})_{ijkl}  =\frac{1}{4}\bigl(\alpha_{ijkl}
+\alpha_{ikjl}-\alpha_{jilk}-\alpha_{jlik}\bigr) \, ,
\\\label{decaj22} &\alpha(j_{2,2}+j'_{2,2})_{ijkl}  =\frac{1}{6}
\bigl(2\alpha_{ijkl}-\alpha_{iklj}-\alpha_{iljk} + 2\alpha_{jilk}
-\alpha_{jkil}-\alpha_{kijl}\bigr) \, ,
\\\label{decaj22p} &\alpha(j_{2,2}-j'_{2,2})_{ijkl}  = \frac{1}{6}
\bigl(-2\alpha_{ijlk} + \alpha_{ikjl} + \alpha_{ilkj}
-2\alpha_{jikl}+\alpha_{jlik}+\alpha_{kjil}\bigr) \, ,
\\\label{decaj211} &\alpha(j_{2,1,1})_{ijkl}  =\frac{1}{4}\bigl(
\alpha_{ijkl}-\alpha_{ikjl}-\alpha_{jilk}+\alpha_{jlik}\bigr) \, ,
\\\label{decaj1111} &\alpha(j_{1,1,1,1})_{ijkl} = \alpha_{[ijkl]}\, .
\end{align}
The consistency conditions fix four of the six irreducible tensors as
\begin{align}\label{solaj4} &\alpha(j_{4})_{ijkl}=\partial_{ijkl}\alpha\, ,\\\label{solaj31}
&\alpha(j_{3,1})_{ijkl} = \frac{2}{3}\bigl(\partial_{(i}\a_{j)[kl]}+
\partial_{(l}\a_{k)[ji]}+
\partial_{(i}\alpha_{k)[jl]}+\partial_{(l}\a_{j)[ki]} \bigr)\, ,
\\\label{solaj22p}
&\alpha(j'_{2,2})_{ijkl} = \frac{2}{3}\bigl(
\partial_{[i}\alpha_{k][lj]}+
\partial_{[j}\a_{l][ki]}+\partial_{[i}\a_{l][kj]}+
\partial_{[j}\a_{k][li]}\bigr)\, ,
\\\label{solaj211}&\alpha(j_{2,1,1})_{ijkl}=
\partial_{i}\a_{l[kj]}+\partial_{j}\a_{k[il]}+\partial_{k}\a_{j[li]}+\partial_{l}\a_{i[jk]}\, .
\end{align}
The tensors $\alpha(j_{2,2})_{ijkl}$ and $\alpha_{[ijkl]}$ are unconstrained. If we define
\be\label{rj22rel} r_{ijkl} = \a(j_{2,2})_{ijkl}=
\frac{1}{6}\bigl( 2\a_{[ij][kl]}+2\a_{[kl][ij]}
-\a_{[ik][lj]}-\a_{[lj][ik]}-\a_{[il][jk]}-\a_{[jk][il]}\bigr)\, ,\ee
then one can show that the irreducible tensor $r$ has precisely the same symmetries as the Riemann curvature tensor,
\be\label{rsym} r_{ijkl}=-r_{jikl}=r_{klij}\, ,\quad r_{ijkl}+r_{iklj}+
r_{iljk}=0\, .\ee
Moreover, the four-form
\be\label{adef} a_{ijkl}=\a_{[ijkl]}=\a(j_{1,1,1,1})_{ijkl}\ee
is simply the unconstrained $A^{(4)}$ in the notation \eqref{AnBndef}.

For the imaginary part coefficients, the condition $\b_{ijkl}=-\b_{lkji}$ let four irreducible tensors,
\begin{align}\label{decbj31}
\begin{split}
&\beta(j_{3,1})_{ijkl} = 
\frac{1}{20}\bigl(5\b_{ijkl}+\b_{ijlk}-4\b_{ikjl}-2\b_{iklj}
\\&\hskip 3cm -2\b_{iljk}+4\b_{ilkj}+\b_{jikl}-3\b_{jilk}-2\b_{jkil}-2\b_{kijl}+4\b_{kjil}\bigr)\, ,
\end{split}
\\\label{decbj31p}
\begin{split}
&\beta(j'_{3,1})_{ijkl} = 
\frac{1}{20}\bigl(5\b_{ijkl}+4\b_{ijlk}+4\b_{ikjl}+2\b_{iklj}
\\&\hskip 3cm +2\b_{iljk}+\b_{ilkj}+4\b_{jikl}+3\b_{jilk}+2\b_{jkil}
+2\b_{kijl}+\b_{kjil}\bigr)
\end{split}\\\label{decbj211}&\b(j_{2,1,1})_{ijkl}= \b_{[ij][kl]}\, ,\\
\label{decbj211p}&\b(j'_{2,1,1})_{ijkl}= \frac{1}{4}\bigl(
\b_{ijkl}-\b_{ilkj}-\b_{jilk}-\b_{kjil}\bigr)\, ,
\end{align}
two of which are fixed by the consistency conditions,
\begin{align}\label{solbj31p}
&\b(j'_{3,1})=\frac{3}{10}\bigl(\partial_{ij}\b_{[kl]}
+2\partial_{ik}\b_{[jl]}+\partial_{il}\b_{[jk]} + 3\partial_{jk}\b_{[il]}
+2\partial_{jl}\b_{[ik]}+\partial_{kl}\b_{[ij]}\bigr)\, ,\\
\label{solbj211p}&\b(j'_{2,1,1}) = \frac{3}{2}\bigl(\partial_{ij}\b_{[kl]}
+\partial_{il}\b_{[jk]}+\partial_{jk}\b_{[li]}+\partial_{kl}\b_{[ij]}
\bigr)\, .
\end{align}
The unconstrained pieces $\b(j_{3,1})$ and $\b(j_{2,1,1})$ can be most easily described in terms of irreducible tensors $s$ and $t$ which are characterized by the following symmetry properties:
\begin{align}\label{ssym} s_{ijkl}& =s_{jikl}=-s_{klij}\, ,\\
\label{tsym} t_{ijkl}&=-t_{jikl}=-t_{klij}\, .
\end{align}
One has
\be\label{sdefirrd}
s_{ijkl}= \frac{1}{2}\bigl(\b(j_{3,1})_{(ij)(kl)}-\b(j_{3,1})_{(kl)(ij)}\bigr)=\frac{1}{5}\tilde\a_{(ij)(kl)}\ee
where
\be\label{atildedef}\tilde\a_{ijkl}=\a_{ijkl}-2\a_{ikjl}\, ,\ee
or equivalently,
\be\label{sj31rel} \b(j_{3,1})_{ijkl}= s_{ijkl}-2s_{ikjl}\, ,\ee
whereas
\be\label{tj211rel} t_{ijkl}=\b(j_{2,1,1})_{ijkl}=\frac{1}{2}\bigl(
\b_{[ij][kl]}-\b_{[kl][ij]}\bigr)\, .\ee
For illustrative purposes, we have provided some details on the derivation of equations \eqref{decbj31}, \eqref{decbj31p} and \eqref{solbj31p} in the Appendix, Section \ref{Appsamplesec}.

\subsection{Summary}

Up to the transformations \eqref{innertrans} and \eqref{outertrans}, an element $\gamma$ of
$\GD$ can be parametrized by an ordinary diffeomorphism $\alpha$ and an infinite set of irreducible tensors
\be\label{irredingamma}
\beta(j_{1,1}),\a(j_{2,1}),\a(j_{2,2}),\a(j_{1,1,1,1}),\b(j_{3,1}),\b(j_{2,1,1})\, , \ldots,\ee
that characterize the expansion \eqref{gamexp}, when the consistency conditions \eqref{realc} and \eqref{gamcond} are taken into account. 
Up to order four, this data can be conveniently described in terms of a set of tensors $b,u,a,r,s,t$; $b$ is a two-form, $a$ is a four-form and the symmetry properties of $u$, $r$, $s$ and $t$ are given by \eqref{usym}, \eqref{rsym}, \eqref{ssym} and \eqref{tsym}. 

Introducing again the upper index, we can thus represent $\gamma\in\GD$ as
\be\label{gammaequiv}\gamma\equiv\bigl( \alpha^{i}(x),b^{i}(x),u^{i}(x),a^{i}(x),r^{i}(x),s^{i}(x),t^{i}(x),\ldots\bigr)\, .\ee
The explicit transformation associated with $\gamma$ is
\begin{multline}
\label{XpXexp}X'^{m}(X)=X'^{m}(x\mathbb I+\e) = \alpha^{m}(x)+
\partial_{i}\alpha^{m}\e^{i} + \frac{1}{2}\partial_{ij}\alpha^{m}
\e^{i}\e^{j} + \frac{i}{4}b_{ij}^{m}\bigl[\e^{i},\e^{j}\bigr]\\
+\frac{1}{6}\partial_{ijk}\alpha^{m}\e^{i}\e^{j}\e^{k}+\frac{1}{18}
u_{ijk}^{m}\bigl[\e^{i},[\e^{j},\e^{k}]\bigr] + \frac{i}{4}\bigl(
\partial_{i}b_{jk}^{m}+\partial_{k}b_{ij}^{m}\bigr)\e^{i}\e^{j}\e^{k}
\\ +\frac{1}{24}\Bigl(\partial_{ijkl}\alpha^{m} + \frac{1}{3}
\bigl( \partial_{i}u_{jkl} + \partial_{l}u_{kji} + \partial_{j}u_{ikl}
+\partial_{k}u_{lji}+4\partial_{i}u_{lkj}+4\partial_{l}u_{ijk}\bigr)
\\+\frac{3i}{5}\bigl(\partial_{ik}b_{jl}+\partial_{jk}b_{li}
+\partial_{jl}b_{ik}+3\partial_{ij}b_{kl}+3\partial_{kl}b_{ij}
+3\partial_{il}b_{jk}\bigr)\Bigr)\e^{i}\e^{j}\e^{k}\e^{l}
\\+ \frac{1}{24}a_{ijkl}\e^{[i}\e^{j}\e^{k}\e^{l]}+ \frac{1}{192}
r_{ijkl} \bigl\{[\e^{i},\e^{j}],[\e^{k},\e^{l}]\bigr\} + \frac{i}{192}
t_{ijkl}\bigl[[\e^{i},\e^{j}],[\e^{k},\e^{l}]\bigr]\\
+ \frac{i}{192}s_{ijkl}\Bigl( \bigl\{[\e^{i},\e^{j}],[\e^{k},\e^{l}]\bigr\}
-2 \bigl\{[\e^{i},\e^{k}],[\e^{j},\e^{l}]\bigr\}\Bigr) + \cdots\, ,
\end{multline}
where $\{A,B\}=AB+BA$ and the $\cdots$ represent terms of higher order.

\subsection{The composition law and the inverse element}

Equipped with the explicit description of the elements $\gamma\in\GD$ in terms of irreducible tensors, we can work out formulas for the composition law and the inverse element in terms of these tensors.

The composition law $\gamma'\circ\gamma$ is straightforwardly obtained from the expansion \eqref{gamexp}. If $X'^{i}=\gamma^{i}(X)$, $X''^{i}=\gamma'^{i}(X')$ and
\be\label{ggpexp} X''^{i}=(\gamma'\circ\gamma)^{i}(X) =
(\gamma'\circ\gamma)^{i}(x\mathbb I+\e) = \sum_{n\geq 0}
\frac{1}{n!}(\gamma'\circ\gamma)^{i}_{i_{1}\cdots i_{n}}(x)
\e^{i_{1}}\cdots\e^{i_{n}}\, ,\ee
we find
\begin{multline}\label{complaw} (\gamma'\circ\gamma)^{i}_{i_{1}\cdots i_{n}}(x)
= \\n!\sum_{k=1}^{n}\frac{1}{k!}\,
\gamma'^{i}_{j_{1}\cdots j_{k}}(x')\!\!\!\!\!\sum_{\substack{m_{i}\geq 1\\
m_{1}+\cdots +m_{k}=n}}\!\!\!\!\frac{1}{m_{1}!\cdots m_{k}!}\,
\gamma^{j_{1}}_{i_{1}\cdots i_{m_{1}}}(x)\cdots
\gamma^{j_{k}}_{i_{m_{1}+\cdots +m_{k-1}+1}\cdots i_{n}}(x)
\, .\end{multline}
The simplest example corresponds to linear transformations $\gamma_{L}$,
\be\label{linearT} \gamma_{L}^{i}(X) = L^{i}_{\ j}X^{j}\, ,\quad
L\in\text{GL}(d,\mathbb R)\, ,\ee
for which $(\gamma_{L})^{i}_{j}=L^{i}_{\ j}$ and $(\gamma_{L})^{i}_{i_{1}\cdots i_{n}}=0$ for $n\geq 2$. Equation \eqref{complaw} then yields
\be\label{linearcomplaw} (\gamma_{L}\circ\gamma)^{i}_{i_{1}\cdots i_{n}}
= L^{i}_{\ j}\gamma^{j}_{i_{1}\cdots i_{n}}\, ,\quad
(\gamma\circ\gamma_{L})^{i}_{i_{1}\cdots i_{n}}
=\gamma^{i}_{j_{1}\cdots j_{n}}L^{j_{1}}_{\ i_{1}}\cdots L^{j_{n}}_{\ i_{n}}\, .\ee
For more general transformations, it is important to realize that 
\eqref{complaw} contains a lot of redundant information. Indeed, 
the coefficients $(\gamma'\circ\gamma)^{i}_{i_{1}\cdots i_{n}}$  automatically satisfy all the consistency conditions discussed in Section
\ref{ccongroupsec} if the coefficients $\gamma^{i}_{i_{1}\cdots i_{n}}$ and $\gamma'^{i}_{i_{1}\cdots i_{n}}$ do. For example, the completely symmetrized version of \eqref{complaw} is simply equivalent to the standard composition law for multiple partial derivatives, consistently with \eqref{asym}, and thus contain no information beyond the fact that $(\gamma'\circ\gamma)^{i}(x) = \gamma'^{i}(\gamma(x))$. 

To obtain the non-trivial information
coded in \eqref{complaw}, we can act with suitable idempotents to isolate the irreducible tensors $b$, $u$, etc, in \eqref{gammaequiv}. Denoting $\gamma\equiv (\alpha=\gamma,b[\gamma],u[\gamma],\ldots)$, $x'=\gamma(x)$, $x''=\gamma'(x')$, $\gamma^{p}_{i}=\partial_{i}\gamma^{p}=\partial x'^{p}/\partial x^{i}$, $\gamma^{p}_{ij}=\partial_{ij}\gamma^{p}=\partial^{2}x'^{p}/\partial x^{i}\partial x^{j}$,
$\gamma'^{m}_{p}=\partial x''^{m}/\partial x'^{p}$, etc, we find, for example,
\begin{align}\label{bcomp} &b[\gamma'\circ\gamma]_{ij}^{m}(x) =\gamma'^{m}_{p}(x')b[\gamma]^{p}_{ij}(x) + \gamma_{i}^{p}(x)\gamma_{j}^{q}(x)b[\gamma']_{pq}^{m}(x')\\
\intertext{and}
\label{ucomp} 
\begin{split}&
u[\gamma'\circ\gamma]_{ijk}^{m}(x) = \gamma'^{m}_{p}(x')u[\gamma]^{p}_{ijk}(x) + \gamma_{i}^{p}(x)\gamma_{j}^{q}(x)\gamma_{k}^{r}(x)u[\gamma']^{m}_{pqr}(x')\\&\hskip .5cm
-\frac{3}{2}b[\gamma']^{m}_{pq}(x')\bigl( \gamma_{i}^{p}(x)b[\gamma]_{jk}^{q}(x) -\gamma_{[j}^{p}(x)b[\gamma]_{k]i}^{q}(x)\bigr)
+\frac{3}{2}\gamma'^{m}_{pq}(x')\gamma_{i[j}^{p}(x)\gamma_{k]}^{q}(x)\, .
\end{split}
\end{align}
Formulas for the higher rank tensors are easy to obtain but they are complicated and not particularly illuminating. In the following, we shall only need \eqref{bcomp} and \eqref{ucomp}.

The inverse element $\gamma^{-1}\equiv (\gamma^{-1},\bar b,\bar u,\ldots)$ can be computed from \eqref{complaw} by imposing $(\gamma^{-1}\circ\gamma)^{i}_{i_{1}\cdots i_{n}}=0$ for $n\geq 2$. If $x'=\gamma(x)$, $\gamma^{p}_{i}=\partial x'^{p}/\partial x^{i}$, $\bar\gamma^{i}_{p}=\partial x^{i}/\partial x'^{p}$, we get for example
\begin{align}\label{binv}
&\bar b^{i}_{pq}(x') = -\bar\gamma^{i}_{r}(x')\bar\gamma^{j}_{p}(x')
\bar\gamma^{k}_{q}(x') b^{r}_{jk}(x)\\
\intertext{and}
\label{uinv}
\begin{split} & \bar u^{i}_{pqr}(x') = -\bar\gamma^{i}_{s}(x')
\bar\gamma^{j}_{p}(x')\bar\gamma^{k}_{q}(x')\bar\gamma^{l}_{r}(x')
u^{s}_{jkl}(x)\\
& \hskip .5cm+\frac{3}{2}\gamma^{s}_{j}\bigl(\bar b^{i}_{sp}(x')\bar b^{j}_{qr}(x')
-\bar b^{i}_{s[q}(x')\bar b_{r]p}^{j}(x')\bigr) + \frac{3}{2}
\bar\gamma_{p}^{j}(x')\gamma_{jk}^{s}(x)\bar\gamma^{i}_{s[q}(x')
\bar\gamma^{k}_{r]}(x')\, .
\end{split}
\end{align}
\subsection{The Lie algebra and the adjoint representation}

It is useful to first briefly review the case $k=1$ of ordinary diffeomorphisms. An infinitesimal diffeomorphism $\gamma$ can be written as 
\be\label{infdiffeo}
\gamma^{i}(x) = y^{i}(x) = x^{i}+\xi^{i}(x)\, ,\ee
where $\xi^{i}$ is the infinitesinal generator.
If we change the coordinate system from $x$ to $x'$, the same infinitesimal diffeomorphism $\gamma$ will be expressed as
\be\label{adinfdif} \gamma'^{i}(x')=y'^{i}(x') = x'^{i}+\xi'^{i}(x')\, ,\ee
with 
\be\label{vecTL} \xi'^{i}(x') = \frac{\partial x'^{i}}{\partial x^{j}}\xi^{j}(x)\, .\ee
This shows that the Lie algebra of the diffeomorphism group is identified with the set of vector fields. By definition, the Lie bracket $[\xi_{1},\xi_{2}]$ between two generators $\xi_{1}$ and $\xi_{2}$ of infinitesimal diffeomorphisms $\gamma_{1}$ and $\gamma_{2}$ is the generator of the infinitesimal diffeomorphism $\gamma_{2}\circ\gamma_{1}\circ\gamma_{2}^{-1}\circ\gamma_{1}^{-1}$. A simple calculation then shows that
\be\label{Lieder} [\xi_{1},\xi_{2}]^{i}=\xi_{1}^{j}\partial_{j}\xi_{2}^{i}
-\xi_{2}^{j}\partial_{j}\xi_{1}^{i} = (\mathcal L_{\xi_{1}}\xi_{2})^{i}
= -(\mathcal L_{\xi_{2}}\xi_{1})^{i}\, ,\ee
where $\mathcal L_{\xi}$ is the usual Lie derivative with respect to the vector field $\xi$. In particular, since $[\xi_{1},\xi_{2}]$ is by construction an infinitesimal generator, we know from \eqref{vecTL} that it must transform as a vector field. This simple remark provides a calculation-free proof of the well-know fact that the Lie derivative \eqref{Lieder} of a vector field is indeed a vector field.

The transformation from \eqref{infdiffeo} to \eqref{adinfdif} can be given a slightly different interpretation. Instead of considering a coordinate change, which is a passive transformation in the sense that it does not act on the points of the base manifold and does not change the diffeomorphism $\gamma$, we can consider the adjoint action
of the diffeomorphism group on itself, $\gamma' = \Gamma\circ\gamma\circ\Gamma^{-1}$ for any diffeomorphism $\Gamma$. With this interpretation, the coordinate change is replaced by the active diffeomorphism $\Gamma$, with $x'^{i}=\Gamma^{i}(x)$. The diffeomorphisms $\gamma$ and $\gamma'$ are then distinct and the formula \eqref{vecTL} no longer relates the component of the same vector field in two coordinate systems but instead maps one vector field $\xi$, the generator of $\gamma$, to another vector field $\xi'=\Gamma_{*}\xi$, the generator of $\gamma'$. Of course, the two interpretations, active or passive, are equally valid.

Let us now see how the above standard results generalize to the non-commutative case $k>1$. An element $\gamma$ of $\GD$ is characterized by an ordinary diffeomorphism and 
by an infinite set of irreducible tensors $(b,u,a,r,s,t,\ldots)$, as in \eqref{gammaequiv}. An infinitesimal transformation will thus be parameterized by an infinitesimal vector field $\xi$ together with infinitesimal tensors $(\mathsf b,\mathsf u,\mathsf a, \mathsf r,\mathsf s,\mathsf t,\ldots)$. In other words, an arbitrary element $\Xi$ of the Lie algebra $\GDL$ is identified with a set of tensors,
\be\label{Xiequiv}\Xi\equiv\bigl( \xi,\mathsf b,\mathsf u,\mathsf a,\mathsf r,
\mathsf s,\mathsf t,\ldots\bigr)\, ,\ee
which have exactly the same symmetry properties as the corresponding tensors para\-me\-tri\-zing the elements of $\GD$ themselves. 

The adjoint action of $\GD$ on itself, or on $\GDL$, can be computed straightforwardly. For the simplest linear $\text{GL}(d,\mathbb R)$ transformations \eqref{linearT}, equation \eqref{linearcomplaw} implies that
\be\label{linearad} (\gamma_{L}\circ\gamma\circ\gamma_{L}^{-1})^{i}_{i_{1}\cdots i_{n}} = L^{i}_{\ j}L_{i_{1}}^{\ j_{1}}\cdots L_{i_{n}}^{\ j_{n}}\gamma^{i}_{i_{1}\cdots i_{n}}\, ,\ee
where $L^{i}_{\ k}L_{j}^{\ k}=\delta^{i}_{j}$. This shows that the coefficients $\gamma^{i}_{i_{1}\cdots i_{n}}$ transform as tensors under $\text{GL}(d,\mathbb R)$. Of course, the same is true for the irreducible pieces $(b,u,a,r,s,t,\ldots)$ in \eqref{gammaequiv} or the 
$(\mathsf b,\mathsf u,\mathsf a,\mathsf r,\mathsf s,\mathsf t,\ldots)$ in \eqref{Xiequiv}. This property actually justifies the use of the terminology ``tensor'' for these objects.

The transformation laws under a general $\GD$ transformation are much more complicated and interesting than simple tensor transformation laws. For example, the action of $\Gamma\equiv (\Gamma^{i}(x)=x'^{i}(x),
B(x),\ldots)$ on \eqref{Xiequiv} yields \eqref{vecTL} and
\begin{multline}\label{btL} 
\mathsf b'^{k}_{ij}(x') =
\frac{\partial x^{m}}{\partial x'^{i}}
\frac{\partial x^{n}}{\partial x'^{j}}\,\biggl(
\frac{\partial x'^{k}}{\partial x^{l}}
\mathsf b^{l}_{mn}(x)
\\+
\xi^{r}(x)\partial_{r}B^{k}_{mn}(x)+\partial_{m}\xi^{r}(x)B_{rn}^{k}(x)
+\partial_{n}\xi^{r}(x)B_{mr}^{k}(x) - \partial'_{p}\xi'^{k}(x')B_{mn}^{p}(x)
\biggr)\, .
\end{multline}
The first line in \eqref{btL} is the standard tensor transformation law, whereas the second line represents a new term given in terms of a sort of bi-local Lie derivative of the tensor $B$. The fact that such bi-local terms enter is natural, since the transformation $\Gamma$ really links the points $x$ and $x'=\Gamma(x)$, with the upper indices on the various tensors parametrizing $\Gamma$ being associated with $x'$ and the lower indices being associated with $x$. This bi-locality is actually already visible in the tensor term, which involves both $\partial x'/\partial x$, which is naturally evaluated at $x$, and $\partial x/\partial x'$, which is naturally evaluated at $x'$. Similar transformation laws can be straightforwardly derived for $\mathsf u$ and the other tensors.

More interesting is the computation of the Lie algebra. The Lie algebra is automatically equipped with a bracket which provides a non-com\-mu\-ta\-ti\-ve, $k>1$, generalization of the Lie derivative \eqref{Lieder}. If
\be\label{GenLie} \bigl[\Xi_{1},\Xi_{2}\bigr]
= \bigl[( \xi_{1},\mathsf b_{1},\mathsf u_{1},\ldots),
( \xi_{2},\mathsf b_{2},\mathsf u_{2},\ldots)\bigr] = 
\Xi_{3}=( \xi_{3},\mathsf b_{3},\mathsf u_{3},\ldots)\, ,\ee
we find
\begin{align}\label{commut1} & \xi_{3}=[\xi_{1},\xi_{2}]=
\mathcal L_{\xi_{1}}\xi_{2}=-\mathcal L_{\xi_{2}}\xi_{1}\, ,
\\\label{commut2}
& \mathsf b_{3} = \mathcal L_{\xi_{1}}\mathsf b_{2}-\mathcal L_{\xi_{2}}
\mathsf b_{1}\, ,\\
\label{commut3}
\begin{split}
&(\mathsf u_{3})^{l}_{ijk} = (\mathcal L_{\xi_{1}}\mathsf u_{2})^{l}_{ijk}
-(\mathcal L_{\xi_{2}}\mathsf u_{1})^{l}_{ijk}\\ &\hskip 1cm
+ \frac{3}{2}\Bigl((\mathsf b_{1})^{l}_{m[j} (\mathsf b_{2})^{m}_{k]i}
- (\mathsf b_{2})^{l}_{m[j} (\mathsf b_{1})^{m}_{k]i}
- (\mathsf b_{1})^{l}_{mi} (\mathsf b_{2})^{m}_{jk}
+ (\mathsf b_{2})^{l}_{mi} (\mathsf b_{1})^{m}_{jk}\Bigr)\\
&\hskip 7cm
+\frac{3}{2}\Bigl( \partial_{i[j}\xi_{1}^{m}\partial_{k]m}\xi_{2}^{l}
- \partial_{i[j}\xi_{2}^{m}\partial_{k]m}\xi_{1}^{l}\Bigr)\, ,
\end{split}
\end{align}
and more and more complicated formulas for the higher tensors. 

A particularly interesting property of the generalized Lie bracket is its covariance with respect to the adjoint action, which generalizes in a rather non-trivial way the covariance of the ordinary Lie derivative. 
For example, if 
$\mathsf b_{1}$ and $\mathsf b_{2}$ transform as in \eqref{btL}, then 
$\mathsf b_{3}$ given by \eqref{commut2} must also transform in the same way. If $B=0$, this is the usual notion of covariance, which is manifest in formula \eqref{commut2} from the covariance of the Lie derivative. When $B\not = 0$, we obtain a non-trivial generalization of the notion of covariance,
which can of course be checked explicitly by plugging the transformation laws \eqref{vecTL} and \eqref{btL} on the right-hand side of \eqref{commut2}.

\subsection{\label{liftsec} On the lift of ordinary diffeomorphisms}

Let us now give a simple proof of an interesting result pointed out in \cite{gc}. First, equation \eqref{commut3} has an interesting consequence.

\noindent\textsc{Lemma}: \emph{The set of elements of $\GD$ of the form}
\be\label{ftrans} X'^{i}=\sum_{n\geq 0}\frac{1}{n!}\partial_{i_{1}\cdots
i_{n}}f^{i}(x)\,\e^{i_{1}}\cdots\e^{i_{n}}\, ,\ee
\emph{where $f$ is an ordinary diffeomorphism, does not form a subgroup of $\GD$ if $k\geq 2$.}

\noindent\textsc{Proof}: Let us first note that the transformation \eqref{ftrans} does satisfy all the consistency conditions of Section \ref{ccongroupsec} and thus does belong to $\GD$. In the representation \eqref{gammaequiv}, it has $\alpha^{i}=f^{i}$ and all the tensors $b$, $u$, etc, set to zero. However, the commutator of two such transformations will have $u\not = 0$, because the terms in the third line of \eqref{commut3} are non-zero even when $\mathsf b_{1}=\mathsf b_{2}=\mathsf u_{1}=\mathsf u_{2}=0$ (the other tensors do not enter in the formula for $\mathsf u_{3}$). The same result can be obtained from the composition law \eqref{ucomp}, which shows that the product of two transformations of the form \eqref{ftrans} will have $u\not = 0$.

So we see that the simplest representation \eqref{ftrans} of the usual group of diffeomorphism in the larger group $\GD$ is actually inconsistent. With the machinery we have developed, it is actually very simple to prove the much more general result mentioned in \cite{gc}.

\noindent\textsc{Definition}: \emph{A \emph{lift} of the group of ordinary diffeomorphisms $\text{Diff}$ into $\GD$ is a group morphism $\Phi:\text{Diff}\rightarrow\GD$ such that $\Phi(f)\equiv (f^{i},\ldots)$.
}

\noindent\textsc{Theorem}: \emph{There is no lift of $\text{Diff}$ into $\GD$ for $k\geq 2$}.

\noindent\textsc{Proof}: The simplest proof is obtained by working at the level of the Lie algebra. Let us assume that a lift $\Phi$ does exist. If $\xi$ is the generator of $f$, then the generator $\Xi$ of $\Phi(f)$ must be of the form \eqref{Xiequiv}, with the tensors $\mathsf b$, $\mathsf u$, etc, depending linearly on $\xi$. Equivalently, the infinitesimal coefficients $\gamma^{i}_{i_{1}\cdots i_{n}}$, $n\geq 2$, characterizing $\Phi(f)$ must depend linearly on $\xi$.
This linear dependence is strongly constrained by the tensorial transformation law \eqref{linearad} under the action of $\text{GL}(d,\mathbb R)$: 
$\gamma^{i}_{i_{1}\cdots i_{n}}$, for $i\geq 2$, must be proportional to $\partial_{i_{1}\cdots i_{n}}\xi^{i}$. In particular, it must be completely symmetric in its lower indices. The consistency conditions \eqref{asym} and \eqref{bsym} then imply that $\Phi(f)$ must be a transformation of the form \eqref{ftrans}. We deduce from the lemma that
$\Phi(\text{Diff})$ is not a subgroup of $\GD$, which contradicts the fact that $\Phi$ is a group morphism. We conclude that the lift $\Phi$ cannot exist.

A direct consequence of the above theorem is that \emph{there is no action of the group of diffeomorphisms on the space of matrix coordinates $X^{i}$ that respects the $\text U(k)$ gauge symmetry, the single-trace structure and acts in the usual way on the diagonal matrices.} This result might superficially suggest that there is an inconsistency with diffeomorphism invariance in string theory, but of course this is not so as we explain in the next Section.

\section{\label{gtsec} The gauge transformations and applications}
\subsection{\label{closedopensec} Closed strings gauge symmetries versus emergent gauge symmetries}

The apparent paradox discussed in \ref{liftsec} can actually be solved in two different ways.

One way of thinking is to assume the a priori existence of additional structures on top of the matrix coordinates $X^{i}$. This is quite natural in the traditional point of view on string theory, where the closed string modes are on an equal footing with the open string modes. The space of physical variables on which the group of diffeomorphisms has to act is thus no longer the space of matrices $X^{i}$ alone, but a bigger space including the $X^{i}$ alongside with all the supergravity fields, which we denote collectively by $\Sigma$. On this space of fields acts the full gauge group $G_{\text{SUGRA}}$ of supergravity gauge invariances, which includes the $p$-form gauge invariances on top of the diffeomorphisms. If $f\in G_{\text{SUGRA}}$, let us denote the action by $f\cdot \Sigma = \Sigma_{f}(\Sigma)$. It satisfies the consistency condition
\be\label{Sigactcc} \Sigma_{f_{1}}\circ\Sigma_{f_{2}}=\Sigma_{f_{1}f_{2}}\, .\ee
From the results on Section \ref{liftsec}, we know that $G_{\text{SUGRA}}$ does not act on the space of the $X$s alone. However, this does not prevent us to define an action on $(X,\Sigma)$ of the form 
\be\label{Gsugraact} (X,\Sigma)\mapsto f\cdot(X,\Sigma) = \bigl(X_{f}(X,\Sigma),\Sigma_{f}(\Sigma)\bigr)\, .\ee
The crucial difference with the case discussed in \ref{liftsec} is that the transformation $X_{f}$ of the matrix coordinates is \emph{background dependent} through its explicit dependence on $\Sigma$. The condition for a consistent group action now reads
\be\label{Xsigcc} X_{f_{1}f_{2}}(X,\Sigma) = X_{f_{1}}\bigl( X_{f_{2}}(X,\Sigma), \Sigma_{f_{2}}(\Sigma)\bigr)\, ,\ee
together with \eqref{Sigactcc}, and these conditions can a priori be solved.

Indeed, this point of view was advocated in \cite{gc} and an explicit solution of \eqref{Xsigcc} was constructed up to order four for $f\in\text{Diff}\subset
G_{\text{SUGRA}}$. The transformation $X_{f}$ built in \cite{gc} depends on an arbitrary background metric $g$. It can be expanded as in \eqref{gamexp}, with coefficients $\gn$ depending explicitly on $g$. This expansion must satisfy the conditions explained in \ref{ccongroupsec} and thus can be parameterized as in \eqref{gammaequiv}, with $\alpha^{i}=f^{i}$ and tensors $b$, $u$, etc, depending on $g$. For example, the solution of \cite{gc} yields
\be\label{ansd} \alpha^{m}=f^{m}\, ,\quad b[f,g]^{m}_{ij} = 0\, ,\quad
u[f,g]^{m}_{ijk} = -\frac{3}{2}\partial_{n[j}f^{m}\Gamma_{k]i}^{n}\, ,\ee
where the $\Gamma_{ij}^{k}$ are the Christoffel symbols for the background metric $g$. One can easily check that the above definition of $u$ is consistent with the constraints \eqref{usym}. If $x'=f_{1}(x)$ and $x''=f_{2}(x')$, the composition law \eqref{ucomp} shows that the consistency condition \eqref{Xsigcc} is equivalent to
\begin{multline}\label{explccudiff} u[f_{2}\circ f_{1},g]^{l}_{ijk} (x) = 
\frac{\partial x''^{l}}{\partial x'^{q}} u[f_{1},g]^{q}_{ijk}(x)
+ \frac{\partial x'^{m}}{\partial x^{i}}\frac{\partial x'^{n}}{\partial x^{j}} \frac{\partial x'^{p}}{\partial x^{k}} u[f_{2},f_{1}\cdot g]^{l}_{mnp}(x')\\ + \frac{3}{2}\frac{\partial^{2}x''^{l}}{\partial x'^{m}\partial x'^{n}}\frac{\partial^{2}x'^{m}}{\partial x^{i}\partial x^{[j}}
\frac{\partial x'^{n}}{\partial x^{k]}}\,\cdotp
\end{multline}
This equality can be straightforwardly checked from \eqref{ansd} and 
the well-known transformation properties of the metric and the Christoffel symbol under the action of $f_{1}\in\text{Diff}$.

It is plausible that the conditions \eqref{Xsigcc} can be solved to all orders using appropriate formulas for $a[f,g]$, $r[f,g]$, $s[f,g]$, $t[f,g]$ and the higher tensors in \eqref{gammaequiv}. Unfortunately, the solutions to the consistency conditions will not be unique \cite{gc}. In particular, the metric $g$ is arbitrary and is not clearly identified in terms of the supergravity fields (for instance, it could be the string frame metric, or the Einstein frame metric, or the metric seen by some particular D-brane, etc...). Moreover, there is no reason for the transformation $X_{f}$ to depend on the metric alone and more general possibilities may be found by including a dependence in other supergravity fields. Constraints on $X_{f}$ can be found by imposing diffeomorphism invariance, or more generally invariance under $G_{\text{SUGRA}}$, on a particular D-brane action,
\be\label{diffinv} S\bigl( X_{f}(X,\Sigma),\Sigma_{f}(\Sigma)\bigr)
= S\bigl( X,\Sigma\bigr)\, ,
\ee
if one knows the dependence of $S$ on the supergravity fields $\Sigma$, for example by using Myers' results \cite{Myers}. Even with this additional constraint, the solution is not unique. In \cite{gc}, \eqref{diffinv} was actually used the other way around, to put some constraints on the metric dependence of $S$, turning off all the other possible background fields. This is an interesting approach, since, beyond Myers' formulas, little is known about the general non-abelian D-brane actions in curved space. Unfortunately, but not surprisingly, the procedure is highly ambiguous and cannot fix the form of the action. Moreover, considering only the metric dependence might be misleading, since a fully consistent picture may require the closed string background to be on-shell.

Because of all the above-mentioned difficulties, it may be more fruitful to use a different point of view, which is strongly favored if one interprets the closed string background as emerging from a microscopic, open-string like theory, as in the models studied in \cite{fer1,fmr,fm,fr}. In this point of view, the only natural gauge group is the group $\GD$ discussed in Section \ref{ggsec}. The closed string fields emerge from the coefficients $\cn$ in the expansion \eqref{Sexpdef}. Two sets of fields $\{c_{i_{1}\cdots i_{n}},\, n\geq 0\}$ and $\{c'_{i_{1}\cdots i_{n}},\, n\geq 0\}$ will be physically equivalent if they correspond to the expansion of the same action in two different matrix coordinate systems $X$ and $X'$ related to each other by a $\GD$ transformation,
\be\label{Strans} S'(X') = S(X)\, .\ee

Let us emphasize again the difference between \eqref{diffinv} and \eqref{Strans}.
In equation \eqref{diffinv}, the background supergravity fields are given and one considers transformations under $G_{\text{SUGRA}}$ only. The transformation laws of the coefficients of the action, which are related to the supergravity background fields, are fixed a priori. The existence of a transformation law $X_{f}$ on the matrix coordinates such that \eqref{diffinv} is valid is then required by consistency with the invariance under $G_{\text{SUGRA}}$. On the other hand, equation \eqref{Strans} is \emph{not} a consistency requirement, but the \emph{definition} of the action of the group $\GD$ on the coefficients $\cn$ and thus on the supergravity fields. Since $S$ and $S'$ are physically equivalent, $\GD$ is the group of gauge transformations.

How can we see the usual gauge group $G_{\text{SUGRA}}$ emerge and how is the ``paradox'' discussed in Section \ref{liftsec} solved in this picture? The point is that, even though there is no lift of $G_{\text{SUGRA}}$ into $\GD$, the groups $G_{\text{SUGRA}}$ and $\GD$ can act in the same way on a set of fields. For example, one can define the action of $\gamma\in\GD$ on scalar fields $\phi$ as
\be\label{GDactex1} \gamma\cdot\phi (x) = \phi(\alpha^{-1}(x))\, ,\ee
where $\alpha$ is the ordinary diffeomorphism parametrizing $\gamma$ in \eqref{gammaequiv}. This of course coincides with the usual action of Diff on a scalar field. It is obviously a consistent action of Diff, but it is also a consistent action of $\GD$ because of the form of the composition law in $\GD$,
\be\label{comptrivialGD} (\alpha_{1},\ldots)\circ (\alpha_{2},\ldots) =
(\alpha_{1}\circ\alpha_{2},\ldots)\, .\ee
In other words, even though there is no good group morphism $\Phi: G_{\text{SUGRA}}\rightarrow \GD$ in the sense explained in \ref{liftsec}, there do exist surjective group morphisms $\Psi:\GD\rightarrow G_{\text{SUGRA}}$. If we have an action of $\GD$ for which the kernel of $\Psi$ acts trivially, then we can use $\Psi$ to find a corresponding action of $G_{\text{SUGRA}}$. This is the mechanism by which the usual $G_{\text{SUGRA}}$ transformations can emerge consistently from $\GD$ and the open-string description. A simple explicit example, for the case of the $\text{AdS}_{5}\times\text{S}^{5}$ background studied in \cite{fer1}, will be given in Section \ref{AdSsec}. 

It is also important to realize that, in general, the action of $\GD$ will induce transformation laws that are more general than the standard $G_{\text{SUGRA}}$ gauge transformations. In the rest of this Section, we are going to derive the form of these general transformation laws and discuss some of their consequences.

\subsection{\label{gtgensec}The gauge transformations}

Finding the explicit relation between two sets of fields $\cn$ and $\cn'$ related by a $\GD$ gauge transformation is completely straightforward. The action $S(X)$ is expanded as in \eqref{Sexp}, $S'(X')$ is expanded as
\be\label{Spexpab} S'(X')= S'(x'\mathbb I + \e') = \sum_{n\geq 0}\frac{1}{n!}c'_{i_{1}\cdots i_{n}}(x')
\e'^{i_{1}}\cdots\e'^{i_{n}}\, ,\ee
$X$ and $X'$ are related to each other as in \eqref{gamexp} and we impose the equality \eqref{Strans}. This yields a general relation of the form
\be\label{cactiona} c_{i_{1}\cdots i_{n}}(x) = \frac{1}{n}\sum_{k=1}^{n}
\bar c_{i_{k}\cdots i_{n}i_{1}\cdots i_{k-1}}(x)\ee
where
\begin{multline}\label{actionlaw} \bar c_{i_{1}\cdots i_{n}}(x)
= \\n!\sum_{k=1}^{n}\frac{1}{k!}\,
c'_{j_{1}\cdots j_{k}}(x')\!\!\!\!\!\sum_{\substack{m_{i}\geq 1\\
m_{1}+\cdots +m_{k}=n}}\!\!\!\!\frac{1}{m_{1}!\cdots m_{k}!}\,
\gamma^{j_{1}}_{i_{1}\cdots i_{m_{1}}}(x)\cdots
\gamma^{j_{k}}_{i_{m_{1}+\cdots +m_{k-1}+1}\cdots i_{n}}(x)
\, .\end{multline}

The simplest example corresponds to the case where 
$\gamma=\gamma_{L}\in\text{GL}(d,\mathbb R)$ is a linear transformation, as in \eqref{linearT}. Equation \eqref{actionlaw} then yields the ordinary tensorial transformation law,
\be\label{linearaction} c_{i_{1}\cdots i_{n}}(x) = c'_{j_{1}\cdots j_{n}}(x')L^{j_{1}}_{\ i_{1}}\cdots L^{j_{n}}_{\ i_{n}}\, ,\ee
which actually justifies our use of the term ``tensor'' for the coefficients $\cn$ or their associated irreducible pieces.

For general $\GD$ transformations, the transformation laws are much more involved. Let us note, however, that the formulas \eqref{cactiona} and \eqref{actionlaw} contain a lot of redundant information, since the coefficients $\cn$ and $c'_{i_{1}\cdots i_{n}}$ must satisfy the consistency conditions discussed in Section \ref{Taylorsec}. In particular, if the set of coefficients $\{ \cn\}$ satisfy these conditions, then the set $\{
c'_{i_{1}\cdots i_{n}}\}$ determined by \eqref{cactiona} and \eqref{actionlaw} automatically satisfy these conditions as well, and vice versa. All the information is thus contained in the transformation rules for the independent irreducible tensors \eqref{Sequiv}, expressed in terms of the independent irreducible tensors \eqref{gammaequiv} parametrizing the transformation law itself. 

The calculations required to express the  transformation laws in this way are rather involved. We focus on the tensors $c$, $c_{[ijk]}$, $c_{[ij][kl]}$ and $c_{[ijklm]}$ which, as explained in \ref{Myerssec}, encode Myers' action, and whose transformation laws will be explicitly used in the applications presented in \ref{apppsec} and \ref{AdSsec}. We find 
\begin{align}\label{ctrans} & c(x)=c'(x')\, ,\\\label{c3trans}
&
c_{[ijk]}(x) = \gamma^{m}_{i}\gamma^{n}_{j}\gamma^{p}_{k}
c'_{[mnp]}(x')
+ 3i \partial'_{m}c'(x')\, \partial^{\hphantom{m}}_{[i}\! b^{m}_{jk]}
+ 3i \partial'_{mn}c'(x')\, b^{m}_{[ij}\gamma^{n}_{k]}\, ,
\\\label{c4trans}
\begin{split}
&
c_{[ij][kl]}(x) = \gamma^{m}_{i}\gamma^{n}_{j}\gamma^{p}_{k}
\gamma^{q}_{l}c'_{[mn][pq]}(x')
+\frac{3i}{2} c'_{[mnp]}(x') \Bigl( b^{m}_{ij}\gamma_{k}^{n}\gamma_{l}^{p}
+ b^{m}_{kl}\gamma^{n}_{i}\gamma^{p}_{j} + 2 \gamma^{m}_{[i}b^{n}_{j][k}
\gamma^{p}_{l]}\Bigr)\\ &\hskip 3cm
+\frac{1}{2}\partial'_{m}c'(x')\Bigl(\frac{3}{2} r^{m}_{ijkl} +
\partial^{\hphantom{m}}_{[i}\! u^{m}_{j]kl} + 
\partial^{\hphantom{m}}_{[k}\!u^{m}_{l]ij}\Bigr)
+3\partial'_{mnp}c'(x')\, \gamma^{m}_{[i}\alpha^{n}_{j][k}\gamma^{p}_{l]}
\\&\hskip 1cm
+ 2\partial'_{mn}c'(x')\Bigl( \gamma^{m}_{[i}u^{n}_{j]kl} +
\gamma^{m}_{[k}u^{n}_{l]ij} + \frac{3}{8}\bigl(
\alpha_{jk}^{m}\alpha_{il}^{n}-\alpha_{ik}^{m}\alpha_{jl}^{n}
+b_{jl}^{m}b_{ki}^{n}-b_{jk}^{m}b_{li}^{n}-2 b_{ij}^{m}b_{kl}^{n}
\bigr)\Bigr)\, ,
\end{split}
\\\label{c5trans}
\begin{split}
& c_{[ijklm]}(x) = \gamma^{p}_{i}\gamma^{q}_{j}\gamma^{r}_{k}
\gamma^{s}_{l}\gamma^{t}_{m}c'_{[pqrst]}(x')\\&\hskip 2.5cm
+10i\Bigl( 3 c'_{[pqr]}(x')\,\partial^{\hphantom{p}}_{[i} b^{p}_{jk}
\gamma^{q}_{l}\gamma^{r}_{m]} + \partial'_{p}c'_{[qrs]}(x')\,
b^{p}_{[ij}\gamma^{q}_{k}\gamma^{r}_{l}\gamma^{s}_{m]}\Bigr)\\
&\hskip 2.5cm
+5\partial'_{p}c'(x')\, 
\partial^{\hphantom{p}}_{\raisebox{-0.75pt}{$\scriptstyle{[i}$}}
a^{p}_{jklm]} + 5\, \partial'_{pq}c'(x')\Bigl(
\gamma^{p}_{[i}a^{q}_{jklm]} - 6\, b^{p}_{[ij}
\partial^{\hphantom{q}}_{\raisebox{-1.25pt}{$\scriptstyle{k}$}}
b^{q}_{lm]}\Bigr)\\&\hskip 9cm
-15\,\partial'_{pqr}c'(x')\, b^{p}_{[ij}b^{q}_{kl}\gamma^{r}_{m]}\, .
\end{split}
\end{align}
The transformation rules for $c_{[ijk]}$ and $c_{[ijklm]}$ can be most conveniently rewritten in the form language, using the definitions \eqref{Cndef} and 
\be\label{formlanguage} \d x'^{m}=\gamma^{m}_{i}\d x^{i}, \quad
b^{m}=\frac{1}{2}b^{m}_{ij}\d x^{i}\wedge\d x^{j}\, .\ee
Equations \eqref{c3trans} and \eqref{c5trans} are then equivalent to
\begin{align}\label{F3trans}&
F^{(3)} = F'^{(3)} + i\,\d\Bigl( \partial'_{m}c'\, b^{m}\Bigr)\, ,\\
\label{F5trans} &
F^{(5)} = F'^{(5)} +\d\Bigl( \partial'_{m}c'\, a^{m}-\frac{1}{2}
\partial'_{mn}c'\, b^{m}\wedge b^{n} + \frac{i}{2} c'_{[mnp]}\,
b^{m}\wedge\d x'^{n}\wedge\d x'^{p}\Bigr)\, .
\end{align}
Let us note that standard tensorial transformation laws would correspond to $F^{(3)}=F'^{(3)}$ and $F^{(5)}=F'^{(5)}$. The additional terms enter because of the non-commutative structure of the space of matrix coordinates. The fact that $F'^{(3)}-F^{(3)}$ and $F'^{(5)}-F^{(5)}$ turns out to be exact forms is perfectly consistent with the constraints \eqref{dC}. Similarly, the simple tensorial transformation law of $c_{[ij][kl]}$, which would correspond to the first term on the right-hand side of \eqref{c4trans}, is supplemented by additional terms which, of course, are consistent with the symmetries \eqref{riemsym}.

The form of the gauge transformations \eqref{ctrans}--\eqref{F5trans} are quite interesting and non standard. Their form is, to some extent, dictated by the non-trivial structure of the group $\GD$ discussed in Section \ref{ggsec}. We are now going to provide a few simple applications and clarify their physical meaning.

\subsection{\label{apppsec} Application to $p$-form gauge transformations}

As a first application, let us show how the $p$-form supergravity gauge transformations are generated from the $\GD$ gauge transformations and thus naturally emerge from the open string description. We shall treat below the case of Myers' D-instanton action and in Section \ref{qmsec} the case of D-particles. In particular, we are going to check explicitly the consistency of Myers' action with the $p$-form gauge symmetries via equation \eqref{diffinv}. 

An interesting feature, first derived in \cite{ginv}, is that the $B$-field gauge transformations must act non-trivially on the matrix coordinates, with $\delta X\sim [X,X]$. This means that the transformation $X_{f}$ in \eqref{diffinv} is non-trivial when $f\in G_{\text{SUGRA}}$ corresponds to a $B$-field gauge transformation. This result may be surprising from the closed string perspective but, from the discussion in \ref{closedopensec}, it is perfectly natural from the emergent geometry, or open string, point of view.

The references \cite{ginv} focused on the Chern-Simons part of the action and on the leading order transformation law for the matrix coordinates. As we now discuss, the formalism that we have developed so far allows us to  generalize effortlessly the analysis to the full non-abelian D-brane action, including the Dirac-Born-Infeld part, and to work out the matrix coordinates transformation laws up to the fourth order.

Myers' action was discussed in \ref{Myerssec} and its dependence on the supergravity $p$-forms is coded in the forms $F^{(1)}$, $F^{(3)}$ and $F^{(5)}$ given in \eqref{F1F3F5myers}. The Ramond-Ramond two- and four-forms gauge transformations are parametrized by a one-form $\mu$ and a three-form $\omega$ and induce the following non-trivial variations on the form fields,
\be\label{C2C4gi} \Delta C_{2} = \d\mu\, ,\quad \Delta C_{4}=\d\omega
+ H\wedge\mu\, ,\ee
where $H=\d B$ is the Neveu-Schwarz three-form field strength. It is immediate to check that $F^{(1)}$, $F^{(3)}$ and $F^{(5)}$ do not change under these transformations and thus the D-brane action is trivially invariant. Much more interesting is the case of the $B$-field gauge transformations, which acts only on $B$ as
\be\label{Bgi} \Delta B = \d\lambda\, .\ee
It yields
\begin{align}\label{F1sugragi} \Delta F^{(1)} & = 0\, ,\\
\label{F3sugragi} \Delta F^{(3)} & = -\frac{2i}{\ls^{2}}F^{(1)}\wedge\d\lambda\, ,\\\label{F5sugragi}
\Delta F^{(5)} & = -\frac{6i}{\ls^{2}} F^{(3)}\wedge\d\lambda
-\frac{6}{\ls^{4}} F^{(1)}\wedge\d\lambda\wedge\d\lambda\, .
\end{align}
The quadratic term on the right-hand side of \eqref{F5sugragi} ensures that the composition of two gauge transformations associated with $\lambda_{1}$ and $\lambda_{2}$ yields a gauge transformation of the same type with $\lambda = \lambda_{1}+\lambda_{2}$.

It is straightforward to check that the formulas \eqref{F1sugragi}--\eqref{F5sugragi} are special cases of the general $\GD$ gauge transformations \eqref{F3trans} and \eqref{F5trans}, for $x'=x$ (the associated standard diffeomorphism is trivial, as expected) and
\be\label{absolgi} b^{i}=\frac{2}{\ls^{2}}\,\lambda\wedge\d x^{i}\, ,\quad 
a^{i} = -\frac{6}{\ls^{4}}\lambda\wedge\d\lambda\wedge\d x^{i}\, ,\ee
up to an exact one-form which can always be added to $\lambda$. 

There remains to check that the other supergravity fields do not vary. The condition $\Delta\tau=0$ follows from the first equation in \eqref{c0c3myers}, \eqref{ctrans} and $x'=x$. From \eqref{c4myersbis}, the condition $\Delta G_{ij}=0$ is then equivalent to $\Delta c_{[ij][kl]}=0$. On the other hand, $\Delta c_{[ij][kl]}$ is given by \eqref{c4trans}, in the special case for which $x'=x$ (and thus $\gamma_{i}^{m}=\delta_{i}^{m}$) and, from the first equation in \eqref{absolgi}, 
\be\label{bexplicitgi} b^{m}_{ij} = \frac{4}{\ls^{2}}\lambda^{}_{[i}\delta_{j]}^{m}\, .\ee
On can then immediately check that the term proportional to $c'_{[mnp]}$ on the right-hand side of \eqref{c4trans} automatically vanish. On the other hand, the term proportional to $\partial'_{mn}c'$ can be made to vanish by choosing
\be\label{uexplicitgi} u^{m}_{ijk} = -\frac{9}{\ls^{4}}
\lambda^{}_{i}\lambda^{}_{[j}\delta_{k]}^{m} \, ,\ee
and the term proportional to $\partial'_{m}c'$ can then be made to vanish by choosing
\be\label{rexplicitgi} r^{m}_{ijkl} = \frac{6}{\ls^{4}}\Bigl(
\partial^{}_{[i}\bigl(\lambda^{}_{j]}\lambda^{}_{[k}\delta_{l]}^{m}\bigr)
+ \partial^{}_{[k}\bigl(\lambda^{}_{l]}\lambda^{}_{[i}\delta_{j]}^{m}\bigr)\Bigr)\, .\ee
Of course, the tensors $u^{m}_{ijk}$ and $r^{m}_{ijkl}$ defined in this way satisfy the required symmetry properties \eqref{usym} and \eqref{rsym}.

The transformation law on the matrix coordinates given by \eqref{absolgi}, \eqref{bexplicitgi}, \eqref{uexplicitgi} and \eqref{rexplicitgi} turn out to be \emph{background independent}. It is very natural to expect that a background-independent extension of the transformation law to all orders could be found. Let us also note that the analysis can be performed independently on the detailed form of Myers' action and in particular independently of the small $\ls^{2}$ approximation. Indeed, the background field transformation laws \eqref{F1sugragi}, \eqref{F3sugragi} and \eqref{F5sugragi} are consistent with the general constraints \eqref{dC} discussed in Section \ref{Taylorsec} and thus well-defined for any matrix action.

\subsection{\label{AdSsec} Diffeomorphisms and the emergent $\text{AdS}_{5}\times\text{S}^{5}$ background}

As a consequence of the theorem reviewed in \ref{liftsec}, the discussion of the previous subsection cannot be generalized straightforwardly to the case of space-time diffeomorphisms, because background-independent transformation laws associated with diffeomorphisms do not exist for the matrix coordinates. However, consistency with diffeomorphism invariance can nevertheless be achieved, as explained in \ref{closedopensec}.

Let us see explicitly how this works for the D-instanton action in the presence of D3-branes, which was derived from a microscopic calculation in \cite{fer1}. The action $S(X)$ turns out to be precisely of the form predicted by Myers, as in equations \eqref{c0c3myers}, \eqref{c4myersbis} and \eqref{c5myers}. The axion-dilaton $\tau$ is a constant and is expressed in terms of the $\vartheta$ angle and 't~Hooft coupling $\lambda$ of the $\nn=4$ gauge theory living on the D3 branes as
\be\label{tauN4} \tau = \frac{\vartheta}{2\pi} + \frac{4i\pi N}{\lambda} \, \cdotp\ee
The coefficient $c_{[ij][kl]}$ factorizes as in \eqref{c4myersbis} in terms of the usual Euclidean $\text{AdS}_{5}\times\text{S}^{5}$ metric,
\be\label{AdSmet} \d s^{2} = G_{ij}\d x^{i}\d x^{j} = \frac{r^{2}}{R^{2}}
\d x^{\mu}\d x^{\mu} + \frac{R^{2}}{r^{2}}\d r^{2} + R^{2}\d\Omega_{5}^{2}\, ,\ee
where $1\leq\mu\leq 4$, $\d\Omega_{5}^{2}$ is the metric for the unit round five-sphere and the radius $R$ is given by
\be\label{radiusAdS} R^{4} = \frac{\ls^{4}\lambda}{4\pi^{2}}\, \cdotp\ee
The form coefficients \eqref{F1F3F5myers} are found to be
\be\label{formN4sol} F^{(1)}=0\, ,\quad F^{(3)} = 0\, ,\quad
F^{(5)} = -\frac{96 N}{R^{5}}\bigl(\omega_{\text{AdS}_{5}}+ i \omega_{\text{S}^{5}}\bigr)\, ,\ee
where $\omega_{\text{AdS}_{5}}$ and $\omega_{\text{S}^{5}}$ are the volume forms associated with the $\text{AdS}_{5}$ and $\text{S}^{5}$ factors of the metric \eqref{AdSmet}. Formulas \eqref{tauN4}--\eqref{formN4sol} reproduce precisely the
$\text{AdS}_{5}\times\text{S}^{5}$ background of type IIB supergravity. 
In particular, the condition $F^{(1)}=0$ comes from the fact that the axion-dilaton is constant, $F^{(3)}=0$ is equivalent to $B=C_{2}=0$ and $F^{(5)}$ yields the correct Ramond-Ramond five-form field strength.

The above solution is derived from the microscopic computation of $S(X)$, not from solving the supergravity equations of motion. By construction, it is then only defined modulo the general $\GD$ gauge transformations discussed in \ref{gtgensec}. Because $c$ is constant and $c_{[ijk]}=0$, the complicated transformation laws \eqref{ctrans}--\eqref{c5trans} actually simplify, for example
\be\label{diffeoN4} c_{[ij][kl]}(x) = \gamma_{i}^{m}\gamma_{j}^{n}
\gamma_{k}^{p}\gamma_{l}^{q}c'_{[mn][pq]}(x)\, ,\quad
c_{[ijklm]}(x) = \gamma_{i}^{p}\gamma_{j}^{q}
\gamma_{k}^{r}\gamma_{l}^{s}\gamma_{m}^{t}c'_{[pqrst]}(x)\, ,
\ee
where $\gamma_{i}^{m}=\partial x'^{m}/\partial x^{i}$. We simply find the action of ordinary diffeomorphisms, emerging from the field redefinition redundancy in the open string point of view. This is perfectly in line with the emerging space philosophy and the discussion around equation \eqref{comptrivialGD}. It is also interesting to find that tensorial quantities in ordinary spacetime, like a metric or a five-form, can emerge from a purely scalar function of non-commuting matrix coordinates. The mechanism at work is quite different from the usual coupling of the metric to a kinetic term or of a $p$-form to a $p$-dimensional worldvolume, for instance. 

\subsection{\label{physicssec} Comments on the general case}

The discussion of the previous subsection uses heavily the special properties of the $\text{AdS}_{5}\times\text{S}^{5}$ background. If $c$ is constant, corresponding to a constant axion-dilaton, we could still implement the ordinary diffeomorphisms with a $\GD$ gauge transformation for which $b^{m}=0$, or more generally of the form $b^{m}=\d\varphi\wedge\d x^{m}$, which ensures that the transformation laws \eqref{ctrans}--\eqref{c5trans} reduce to the standard tensor transformation laws. However, as we have emphasized again and again, this is not natural. One should really consider the general action of $\GD$ and draw the general consequences of the associated transformation laws.

Actually, for a generic background, the action of $\GD$ on the action $S(X)$ is very drastic. In the abelian case, $k=1$, this is well-known. If we assume that $\d\re c(x)\wedge\d\im c(x)\not = 0$ then, in the vicinity of $x$, we can always pick a coordinate system such that $x'^{1}=\re c$ and $x'^{2}=\im c$. In this coordinate system, $S'(x') = x'^{1} + i x'^{2}$ is a simple linear function. In the non-commutative case, we would like to make a similar statement.

\noindent\textsc{Claim}: \emph{Let us assume that 
$\d\re c(x)\wedge\d\im c(x)\not = 0$. Then it is always possible to gauge away all the coefficients $\cn(x)$ for $n\geq 2$ by using a general $\GD$ gauge transformation.}

Up to order five, this is straightforwardly proved from our explicit formulas \eqref{ctrans}--\eqref{F5trans} and the elementary

\noindent\textsc{Lemma}: \emph{If $v_{m}$ is a complex valued vector such that $\re v_{m}$ and $\im v_{m}$ are linearly independent and if $\rho_{\iin}$ are arbitrary complex-valued coefficients, then it is always possible to solve the equations}
\be\label{rhogamrel} v_{m}r^{m}_{\iin} = \rho_{\iin}\ee
\emph{for some real coefficients $r^{m}_{\iin}$.}

\noindent Using this lemma, we can choose the a priori arbitrary two- and four-forms $c'_{m} b^{m}$ and $c'_{m} a^{m}$ in \eqref{F3trans} and \eqref{F5trans} in such a way that the closed forms $F'^{(3)}$ and $F'^{(5)}$ vanish. Similarly, the term $\frac{3}{4}c'_{m}r^{m}_{ijkl}$ in \eqref{c4trans} can be adjusted to any tensor with the general symmetries of $c_{[ij][kl]}$ and can thus be used to make $c'_{[mn][pq]}$ vanish.

An all order analysis is beyond the scope of our work, but it is interesting to mention that it is essentially equivalent to the following very natural ``lift'' theorem. We have defined the general notion of a single-trace function $f(X)$ in the beginning of Section \ref{Taylorsec}, via an expansion
\be\label{fdefgen} f(X) =f(x\mathbb I+\e) = \sum_{n\geq 0}\frac{1}{n!}c_{\iin}(x)\tr\e^{i_{1}}\cdots\e^{i_{n}}\, ,\ee
where the cyclic coefficients $c_{\iin}$ must satisfy the constraints \eqref{coefcond2}.  A similar notion of a no-trace matrix-valued function $F(X)$ can be defined as well, via the expansion
\be\label{f0expdef} F(X)=F(x\mathbb I + \e) = \sum_{n\geq 0}\frac{1}{n!}\rho_{\iin}(x)\e^{i_{1}}\cdots\e^{i_{n}}\, ,\ee
where the coefficients $\rho_{\iin}$ satisfy the constraints
\be\label{condontho} \partial_{i}\rho_{\iin} = \bigl(J_{n+1}\cdot\rho\bigr)_{i\iin}\, .\ee
These constraints are the same as in \eqref{gamcond2} and ensure, as usual, the invariance under the shifts \eqref{ccond}. Examples of no-trace matrix-valued functions are the $\gamma^{i}(X)$ defining an element $\gamma\in\GD$ in \eqref{gamexp} or $\hat f(X)$ defined by \eqref{fhatdef}. The lift conjecture then states that \emph{any single-trace function is the trace of a no-trace matrix-valued function}. In other words, given cyclic coefficients satisfying \eqref{coefcond2}, it is always possible to find coefficients $\rho_{\iin}$ satisfying \eqref{condontho} and such that
\be\label{rhocrelation} c_{\iin} = \bigl(\jZn\cdot\rho\bigr)_{\iin}\, .\ee
The action of $\jZn$ is defined in \eqref{jzndef} and takes the cyclic combination of the coefficients $\rho_{\iin}$. This statement seems extremely natural, but the proof is not trivial. For example, it can be easily checked that the choice $\rho_{\iin}=c_{\iin}$ is not consistent.

Assuming this result to be correct, we can then proceed as follow to trivialize $S(X)$. First, we use $\d\re c\wedge\d\im c\not = 0$ to choose the ordinary diffeomorphism $x'(x)$ in $\gamma$ such that 
$x'^{1}=\re c(x)$ and $x'^{2}=\im c(x)$. In other words,
\be\label{cpmcond} c'_{m} = \delta_{m,1}+i \delta_{m,2}\, .\ee
Next, we pick an arbitrary complex-valued tensor $\rho_{\iin}$ satisfying the constraints 
\be\label{condonrho} \partial_{i_{1}}\rho_{i_{2}\cdots i_{n+1}}=
\bigl(J_{n+1}\cdot\rho\bigr)_{i_{1}\cdots i_{n+1}}\, ,\ee
and we choose the coefficients $\gamma^{m}_{\iin}$ for $n\geq 2$ in such a way that
\be\label{rhodef} c'_{m}\gamma^{m}_{\iin}=\rho_{\iin} \, .\ee
This is always possible. Indeed, by taking the real and imaginary parts and using \eqref{cpmcond} on the one hand and \eqref{realcr}, \eqref{realci} on the other hand, we see that \eqref{rhodef} is equivalent to
\begin{align}\label{alpharho} \alpha^{1}_{\iin} & = \frac{1}{2}\re\bigl(
\rho_{\iin}+\rho_{i_{n}\cdots i_{1}}\bigr)\, ,\quad
\alpha^{2}_{\iin}  = \frac{1}{2}\im\bigl(
\rho_{\iin}+\rho_{i_{n}\cdots i_{1}}\bigr)\, ,\\\label{betarho}
\beta^{1}_{\iin} & = \frac{1}{2}\im\bigl(
\rho_{\iin}-\rho_{i_{n}\cdots i_{1}}\bigr)\, ,\quad
\beta^{2}_{\iin}  = \frac{1}{2}\re\bigl(
\rho_{i_{n}\cdots i_{1}}-\rho_{\iin}\bigr)\, .
\end{align}
This is consistent, because the only conditions on the coefficients $\alpha^{m}_{\iin}$ and $\beta^{m}_{\iin}$, which are the constraints \eqref{acond} and \eqref{bcond}, are automatically satisfied if \eqref{condonrho} is satisfied, as can be checked straightforwardly.
Now, using the expansions \eqref{Sexpdef} and \eqref{gamexp}, the condition
\be\label{conditiontrivial} S(X)=S'(X')= c'(x') + c'_{m}\tr\e'^{m}\ee
is equivalent to \eqref{rhocrelation}, which can be solved by the lift theorem.

The above discussion is just the beginning of what could be a much more elaborate mathematical study of single-trace functions modulo the action of $\GD$. This study would correspond to an important generalization of the standard singularity theory of ordinary functions \cite{Arnold}, which deals with the classification of the possible expansions around a point modulo the action of diffeomorphisms (or biholomorphisms in the complex case). In view of the many connexions between D-brane physics, single-trace actions and (super)potentials, (singular) Calabi-Yau spaces and matrix models (see e.g.\ \cite{ferMM} and references therein),  
we believe that the development of this theory could have far-reaching consequences.

\section{\label{qmsec} Matrix quantum mechanics}

The analysis of the previous Sections can be straightforwardly generalized to higher dimensional actions. We are going to discuss briefly the case of quantum mechanical single-trace actions, which is used in particular in \cite{fm}. We continue to work in Euclidean signature, if not explicitly stated otherwise, for consistency with the rest of the paper.

\subsection{\label{TaylorQM} The Taylor expansion}
\subsubsection{\label{TGenQM} General discussion}

In the commutative $k=1$ case, it is always possible to choose a gauge in which the time coordinate $x^{d}$ is identified with the parameter $\lambda$ along the worldline. In the general case $k>1$, we assume that such a static gauge still makes sense and set
\be\label{staticgen} X^{d} = \lambda\, .\ee
The D-brane actions derived from string theory are naturally found in this gauge. The quantum mechanical actions we consider,
\be\label{lagrangian} S = \int L\, \d\lambda\, ,\ee
are thus functionals of matrix worldlines given, in parametric form, by $d-1$ matrix coordinate functions $X^{i}(\lambda)$, $1\leq i\leq d-1$. 

The Lagrangian $L$ is assumed to be a single-trace function of $X^{i}(\lambda)$ and its derivatives. We can expand
\be\label{Lderexp} L = \sum_{p\geq 0} L^{(p)}\, ,\ee
where the term $L^{(p)}$ contains $p$ derivatives of the coordinates.
We shall limit ourselves to the two-derivative action, $p\leq 2$, and study the expansion around arbitrary diagonal time-independent configurations,
\be\label{Xexpdef2} X^{i} = x^{i}\mathbb I + \e^{i}(\lambda)\, .\ee
The potential term, $p=0$, is like an ordinary single-trace function \eqref{Sexpdef},
\be\label{order0exp} L^{(0)} = \sum_{n\geq 0}\frac{1}{n!}c_{\iin}^{(0)}(\lambda, x)
\tr\e^{i_{1}}\cdots\e^{i_{n}}\, ,\ee
with $c^{(0)}_{\iin} = c^{(0)}_{i_{n}i_{1}\cdots i_{n-1}}$. As for the one-derivative term, it can be written, using the cyclicity of the trace, as
\be\label{order1exp} L^{(1)} = \sum_{n\geq 0}\frac{1}{n!}
c^{(1)}_{\iin; k} (\lambda, x) \tr\e^{i_{1}}\cdots\e^{i_{n}} \dot\e^{k}\, ,\ee
where
\be\label{covderdef} \dot\e^{k} = \frac{\d\e^{k}}{\d\lambda} + i \bigl[z,\e^{k}\bigr]\ee
is the covariant derivative along the worldline and $z$ the worldline gauge potential. The two-derivative Lagrangian contains only $\e$ and $\dot\e$, up to the addition of total derivative terms. 
To the fourth order, we can arrange the terms as
\be\label{order2exp} L^{(2)} = c^{(2)}_{kl}\tr\dot\e^{k}\dot\e^{l}
+ c^{(2)}_{i;kl}\tr\e^{i}\dot\e^{k}\dot\e^{l} + c^{(2)}_{ij;kl}
\tr\e^{i}\e^{j}\dot\e^{k}\dot\e^{l} + \tilde c^{(2)}_{ij;kl}
\tr\e^{i}\dot\e^{k}\e^{j}\dot\e^{l} + O(\e^{5})\, .\ee

We now impose the invariance under the shift symmetry \eqref{ccond}. Since the shift parameter $a^{i}$ is $\lambda$-independent, this yields constraints on each term $L^{(p)}$ independently of each other. The constraints on the potential term $L^{(0)}$ are of course exactly the same as the ones studied in Section \ref{Taylorsec}. The coefficients $c^{(0)}_{\iin}$ are thus characterized by irreducible tensors as in \eqref{Sequiv}. The constraints on the expansion \eqref{order1exp} match the ones studied in Section \ref{ggsec} for the expansion \eqref{gamexp}, see in particular \eqref{gammaequiv}, since the $\smash{c^{(1)}_{\iin}}$ do not satisfy any cyclicity condition. For example, 
\be\label{c1constQM} c^{(1)}_{i;k}=\partial^{}_{i}c^{(1)}_{k}\, ,\quad 
c^{(1)}_{(ij);k} = \partial^{}_{ij}c^{(1)}_{k}\, ,\ee
whereas $c^{(1)}_{[ij];k}$ is unconstrained. 
We may wish to impose an additional reality condition,
\be\label{realcondc1} \bigl(c^{(1)}_{\iin;k}\bigr)^{*} = c^{(1)}_{i_{n}\cdots i_{1};k}\, ,\ee
if we work in the Minkowskian. 
The coefficients $\smash{c^{(1)}_{k}}$ and $\smash{c^{(1)}_{[ij];k}}$ must then be real and purely imaginary respectively. A similar analysis can be performed on the second derivative Lagrangian \eqref{order2exp}. For example, we find that
\be\label{twodersol} c^{(2)}_{i;(kl)} = \partial_{i}^{} c^{(2)}_{kl}\, ,\quad
\partial_{i}^{}c^{(2)}_{j;kl} = 2 c^{(2)}_{(ij);kl} + \tilde c^{(2)}_{ij;lk}
+\tilde c^{(2)}_{ji;kl}\, ,\ee
and other constraints of a similar type.

\subsubsection{\label{MyersQM} The example of Myers D-particle action}

A particularly interesting example is the D0-brane action in type IIA string theory. In the Euclidean, a single D0-brane has a Lagrangian of the form
\be\label{D01} L = \frac{\sqrt{2\pi}}{\ls}\Bigl[ e^{-\phi}\sqrt{G_{\mu\nu}\dot x^{\mu}\dot x^{\nu}} + i A_{\mu}\dot x^{\mu}\Bigr]\, ,\ee
where $\phi$, $G_{\mu\nu}$ and $A_{\mu}$ are the dilaton, the string-frame metric and the Ramond-Ramond one-form respectively. This is the action for an ordinary charged particule of equal mass and charge $m=q=\sqrt{2\pi}/\ls$ moving in the $d=10$ dimensional metric
\be\label{defg} g_{\mu\nu} = e^{-2\phi} G_{\mu\nu}\, .\ee
Going to the static gauge $\eqref{staticgen}$, $x^{d}=x^{10}=\lambda$, and expanding as in \eqref{Lderexp}, yield
\be\label{LexpcomD0} L^{(0)} + L^{(1)} + L^{(2)} = \frac{\sqrt{2\pi}}{\ls}
\Bigl[i \mathscr A_{\mu}\dot x^{\mu} + \frac{1}{2}
\mathscr H_{ij} \dot x^{i}\dot x^{j}\Bigr]\, ,\ee
with $1\leq i,j\leq 9$, $d=10$ and
\be\label{AHdef}
\mathscr A_{\mu}  = A_{\mu} - i \, \frac{g_{d\mu}}{\sqrt{g_{dd}}}\, \cvp\quad
\mathscr H_{ij}  = \sqrt{g_{dd}}\biggl( \frac{g_{ij}}{g_{dd}} -
\frac{g_{di} g_{dj}}{g_{dd}^{2}}\biggr)\, .\ee
The non-abelian version of this action, valid for an arbitrary number $k\geq 1$ of D-particles, can be computed from Myers' formulas \cite{Myers}. The Dirac-Born-Infeld part of Myers Lagrangian reads
\be\label{LBI} L_{\text{DBI}} = \frac{\sqrt{2\pi}}{\ls}
\Str e^{-\phi}\sqrt{\det Q^{i}_{\ j}} \sqrt{\Bigl[ G_{\mu\nu} + 
E_{\mu i}\bigl((Q^{-1})^{i}_{\ k} - \delta^{i}_{k}\bigr) E^{kj}E_{j\nu}\Bigr]
\dot\e^{\mu}\dot \e^{\nu}}\, ,
\ee
with
\begin{align}\label{Edef} E_{\mu\nu} & = G_{\mu\nu} + B_{\mu\nu}\, ,\\
\label{Qdef} Q^{i}_{\ j} & = \delta^{i}_{j} + i\ls^{-2}[\e^{i},\e^{k}] E_{kj}
\, .
\end{align}
The latin indices always run from $1$ to $9$ whereas the greek indices run from 1 to 10. In particular, $\dot\e^{10}=1$ because of \eqref{staticgen}. The determinant in \eqref{LBI} acts on the indices $i,j$ and not on the $\text{U}(k)$ indices of the matrices $\e$. The $\Str$ is the symmetrized trace on the $\text{U}(k)$ indices defined in \cite{Myers}. It provides the correct ordering up to order five in the expansion in powers of $\e$ but not beyond. The Chern-Simons part of the Lagrangian is given by
\be\label{LCS} L_{\text{CS}} = i\frac{\sqrt{2\pi}}{\ls}\Str \text{P}\Bigl[
e^{i\ls^{-2}i_{\e}i_{\e}}\sum_{q\geq 0} C_{2q +1}\wedge e^{B}\vert_{1-\text{form}}\Bigr]\, .\ee
The $C_{2q +1}$ are the type IIA Ramond-Ramond forms, $i_{\e}$ is the inner product and we keep only the one-form part of the expression in the bracket. The P denotes the $\text{U}(k)$-covariant pull-back to the D-particle worldline,
\be\label{pulldef} P\bigl[ \omega_{\mu}\d x^{\mu}\bigr] = \omega_{d} + \omega_{i}\dot\e^{i}\, .\ee
A rather tedious calculation then yields explicit expressions for the various irreducible tensors parameterizing the Lagrangian. For example, by noting
\be\label{ACdef} A=C_{1}\, ,\quad C=C_{3}\, ,\quad \tilde C = C_{5}\ee
the Ramond-Ramond one-, three- and five-form potentials respectively, we get
\begin{align}\label{potD0a} c^{(0)} & = i\frac{\sqrt{2\pi}}{\ls}
\mathscr A_{d}\, ,\\\label{potD0b}
c^{(0)}_{[ijk]} & = \frac{3\sqrt{2\pi}}{2\ls^{3}}\partial^{}_{[i}
\bigl( C + \mathscr A\wedge B\bigr)_{jk]d}\, ,\\\label{potD0c}
\begin{split}
c^{(0)}_{[ij][kl]} & = -\frac{9\sqrt{2\pi}}{\ls^{5}}\Bigl[
g_{dd}^{3/2}e^{4\phi}\bigl( \mathscr H_{ik}\mathscr H_{jl} - \mathscr H_{jk}\mathscr H_{il}\bigr)\\ &
\hskip 2.5cm
+ \mathscr H_{ik}B_{jd}B_{ld} - \mathscr H_{jk}B_{id}B_{ld} - \mathscr H_{il}B_{jd}B_{kd}
+ \mathscr H_{jl}B_{id}B_{kd}\Bigr]\, ,
\end{split}\\\label{potD0d}
\begin{split}
c^{(0)}_{[ijklm]} & = -\frac{60i\sqrt{2\pi}}{\ls^{5}}\partial^{}_{[i}
\bigl( \tilde C_{jklm]d} + 4 C_{jkl}B_{m]d} + 6 C_{jkd}B_{lm]}\\
&\hskip 6cm + 
3\mathscr A_{d}B_{jk}B_{lm]} + 12 \mathscr A_{j}B_{kd}B_{lm]}\bigr)\, .
\end{split}
\end{align}
Let us note that $C$ and $\tilde C$ are not independent, since the associated field strengths are dual to each other. Explicitly, if $H=\d B$ as usual, we have, in the Euclidean, 
\be\label{duality} F_{4} = \d C + H\wedge A\, ,\quad F_{6} = \d\tilde C
+ H\wedge C = i*F_{4}\, .\ee
Similarly, the first independent coefficients in the one- and two-derivative terms are given by
\begin{align}\label{pot1D0a} c^{(1)}_{k} & = i\frac{\sqrt{2\pi}}{\ls}
\mathscr A_{k}\, ,\\\label{pot1D0b} c^{(1)}_{[ij];k} & =-\frac{2i\sqrt{2\pi}}{\ls^{3}}\Bigl( 2 \mathscr H_{k[i}B_{j]d} + i\bigl( C + \mathscr A\wedge B\bigr)_{ijk}
\Bigr)\, ,\\\label{pot2D0a} c^{(2)}_{kl} & = \frac{\sqrt{2\pi}}{2\ls} \mathscr H_{kl}
\, ,\quad c^{(2)}_{i;[kl]}  = 0\, , \quad \text{etc...}
\end{align}
We have checked explicitly the consistency of the above formulas with the D-particle Lagrangian obtained from the D-instanton action discussed in Section \ref{Myerssec} by performing a T-duality in the direction of $x^{10}$. This method is actually quite efficient. For example, the fifth order coefficient \eqref{potD0d} in the potential is obtained more easily from T-duality than from the explicit Myers action \eqref{LBI} and \eqref{LCS}.
The formulas \eqref{potD0a}--\eqref{pot2D0a} allow to read off the type IIA supergravity background from the D-particle Lagrangian. They are crucially used in \cite{fm} to derive the emergent supergravity background generated by a large number of D4-branes.

\subsection{\label{GTQM} Gauge symmetries}

Let us now discuss the action of the gauge group $\GD$. The general qualitative discussion of Section \ref{closedopensec} applies to the present quantum mechanical case as well.
Since we are in the static gauge \eqref{staticgen}, we limit ourselves to transformations acting on the transverse matrix coordinates $X^{i}$, $1\leq i\leq 9$. Moreover, if we assume that the coefficients appearing in the expansion \eqref{gamexp} do not depend on $x^{d}=\lambda$, then the general gauge transformations for the coefficients $\smash{c^{(0)}_{\iin}}$ of the potential term 
are given by equations \eqref{ctrans}--\eqref{F5trans}. Similarly, we can find the gauge transformations of the fields appearing in the higher derivative terms. For example, 
\begin{align}\label{c10trans} & c^{(1)}_{k}(x) = \gamma_{k}^{m}c'^{(1)}_{m}(x')\, ,\\\label{c12trans}
& c^{(1)}_{[ij];k}(x) = \gamma_{i}^{m}\gamma_{j}^{n}\gamma_{k}^{p}c'^{(1)}_{[mn];p}(x') + i c'^{(1)}_{p;q}(x')
\bigl( 2 \gamma^{p}_{[i} b_{j]k}^{q} + \gamma_{k}^{q}
b_{ij}^{p}\bigr) + 3i c'^{(1)}_{l}(x') \partial^{}_{[i} b^{l}_{jk]}\, ,
\\
\label{c20trans} & c^{(2)}_{kl}(x) = \gamma_{k}^{m}\gamma_{l}^{n}c'^{(2)}_{mn}(x')\, ,\\\label{c22trans} & c^{(2)}_{i;[kl]}(x) = 
\gamma_{i}^{p}\gamma_{k}^{m}\gamma_{l}^{n}c'^{(2)}_{p;[mn]}(x')
+2 ic'^{(2)}_{mn}(x')\gamma_{[k}^{m}b_{l]i}^{n} \, .
\end{align}

As a simple application, we can study the $p$-form supergravity gauge transformations and extend the results of \cite{ginv} to the full Dirac-Born-Infeld plus Chern-Simons non-abelian D-particle action. 
The Ramond-Ramond one- and three-forms gauge transformations are parameterized by a function $h$ and a two-form $\kappa$, with
\be\label{C1C3gi} \Delta A = \d h\, ,\quad \Delta C = \d\kappa + h\,  \d B\, .\ee
If the background fields and the gauge transformation parameters $h$ and $\kappa$ do not depend on $x^{d}$ and $\kappa$ has components on the transverse coordinates $x^{i}$ only, then the invariance of the D-particle action follows from the invariance of the D-instanton action proven in Section \ref{apppsec} and T-duality. In the general case of time-dependent background and general gauge transformations, we have checked, up to order four in the $\e$ expansion, that the Myers' Lagrangian transforms as a total time derivative and thus that the action is invariant, as required.

The $B$-field gauge transformation $\delta B=\d\lambda$
is more interesting, because, as in Section \ref{apppsec}, it must act on the spacetime matrix coordinates in the non-commutative, $k\geq 2$, case. Again, if the background and the gauge transformation parameter $\lambda$ do not depend on $x^{d}$, and if $\lambda$ has components on the transverse coordinates $x^{i}$ only,
then consistency follows from the D-instanton case studied in \ref{apppsec} and T-duality. If we drop this assumption, then the required 
$\GD$ transformations, acting on the transverse matrix coordinates $X^{i}$, will have to depend on the time $x^{d}$ explicitly. The transformation rules \eqref{ctrans}--\eqref{c5trans} and \eqref{c10trans}--\eqref{c22trans} are then generalized, because terms with different number of derivatives in the expansion \eqref{Lderexp} mix under time-dependent $\GD$ transformations. 
Limiting our analysis to the third order in $\e$, we need the 
following generalizations of \eqref{c3trans} (or \eqref{F3trans}) and \eqref{c12trans}, for a $\GD$ transformation associated with the trivial diffeomorphism $x'=x$,
\begin{align}\label{c3newt} F^{(3)} &= F'^{(3)} + i\,\d\bigl (\partial_{m}
c^{(0)}\, b^{m} + c^{(1)}_{m}\,\partial_{d}b^{m}\bigr)\, ,\\
\label{c12newt} 
  c^{(1)}_{[ij];k} &= c'^{(1)}_{[ij];k} + i\bigl(
 2 b_{k[i}^{l}\partial_{j]}^{}c^{(1)}_{l} + b_{ij}^{l}\partial_{l}c^{(1)}_{k}\bigr) + 3i c^{(1)}_{l} \partial^{}_{[i} b^{l}_{jk]} + 2i
 c^{(2)}_{kl}\partial_{d}b_{ij}^{l} \, .
\end{align}
The other relevant transformation laws are unchanged. As in Section \ref{apppsec}, we have to choose the transformation of the coordinates $X^{i}$ such that
\be\label{bQM}
b^{i} = \frac{2}{\ls^{2}}\lambda\wedge\d x^{i}\, .\ee
This goes a long way in generating the $B$-field gauge transformation, but we also have to take into account the T-dual version of the non-trivial transformation law of the matrix coordinate $X^{d}$, supplemented by an additional term when  $\lambda_{d}\not = 0$. The T-dual of $X^{d}$ is the worldline gauge field $z$ and it can be checked that the correct transformation law is given by
\be\label{delz1} \delta z = \frac{1}{\ls^{2}}\bigl( \lambda_{i}\dot\e^{i}+\partial_{i}\lambda_{d}\,\e^{i}\bigr)\, .\ee
The effect of the background-independent field redefinitions associated with \eqref{bQM} and \eqref{delz1} precisely match the effect of the supergravity gauge transformation $\delta B=\d\lambda$.

\section{Acknowledgments}

I would like to thank Antonin Rovai and Micha Moskovic for many useful discussions.
This work is supported in part by the belgian FRFC (grant 2.4655.07) and IISN (grant 4.4511.06 and 4.4514.08).

\vfill\eject

\begin{appendix}
\section{A short review on tensor symmetries}

This Appendix is devoted to a very brief review on the classification of tensor symmetries. This yields very useful calculational techniques that we have implemented in Mathematica and used to perform most of the calculations presented in the main text.

\subsection{Generalities}

We consider the Hilbert space $T_{d}^{n}$ of complex tensors of rank $n$ in $d$ dimensions, with the norm
\be\label{normTn} ||t|| = \sum_{1\leq i_{1},\ldots,i_{n}\leq d}|t_{i_{1}\cdots i_{n}}|^{2}\, .\ee
The symmetric group $\Sn$ acts on $T^{n}_{d}$ in the usual way,
\be\label{SnactTn} (\sigma\cdot t)_{i_{1}\cdots i_{n}} = 
t_{i_{\sigma(1)}\cdots i_{\sigma(n)}}=\tilde\sigma(t)_{i_{1}\cdots i_{n}}\, ,\ee
where $\tilde\sigma$ is the linear operator associated with the permutation $\sigma$.
The \emph{group algebra} $\CSn$, defined to be the set of formal complex linear combinations of the elements of $\Sn$, also acts on $T^{n}_{d}$, by extending the action \eqref{SnactTn} by linearity. The group algebra is endowed with a Hilbert space structure, for which the $n!$ elements of $\Sn$ form an orthonormal basis. It is also convenient to associate, to each element $x\in\CSn$, a linear operator $\hat x$ acting on $\CSn$ by left multiplication, $\hat x(y) = xy$. If $x\in\Sn$, then both $\hat x$ (which acts on $\CSn$) and $\tilde x$ (which acts on $T^{n}_{d}$) are unitary. Moreover, the Hermitian conjugates $\hat x^{\dagger}$ and $\tilde x^{\dagger}$ are both associated with the same element $x^{\dagger}$ of $\CSn$. Explicitly, $x$ and $x^{\dagger}$ can be expanded as
\be\label{xdagger} x = \sum_{\sigma\in\Sn}x_{\sigma}\sigma\, ,\quad
x^{\dagger} = \sum_{\sigma\in\Sn}x_{\sigma}^{*}\sigma^{-1}\, .\ee

A \emph{right-ideal} of $\CSn$, or simply an \emph{ideal}, is a subspace of $\CSn$ stable under right multiplication (there is also a similar notion of left ideals, but for our purposes right ideals are more natural). An ideal $I$ is called \emph{minimal} if any ideal $J\subset I$ is either the trivial ideal $J=\{0\}$ or equal to $I$. It is known that $I$ is minimal if and only if the representation of $\Sn$ induced on $I$ by the right multiplication is irreducible. The same irreducible representation can be associated with distinct ideals; a given irreducible representation of dimension $\delta$ actually occurs with multiplicity $\delta$ in $\CSn$. If $I$ and $I'$ correspond to inequivalent representations, then $xx'=0$ if $x\in I$ and $x'\in I'$. 

The ideals of $\CSn$ have a simple explicit description. An \emph{idempotent} is an element $j\in\CSn$ such that $j^{2}=j$. Then any ideal is of the form
\be\label{idealgen} I = I(j) = \bigl\{ jx\, ,\ x\in\CSn\bigr\}\, .\ee
The idempotent generating a given ideal is not unique. One can easily show that $I(j)=I(j')$ if and only if $j'=j-jx+jxj$ for some $x\in\CSn$. However, there is a unique Hermitian generating idempotent $j_{I}=j_{I}^{\dagger}$. This idempotent corresponds to the orthogonal projection of the identity element on the ideal $I$ and as such it can be easily constructed algorithmically. The operator $\hat\jmath_{I}$ is the orthogonal projector on $I$.

Minimal ideals are generated by \emph{primitive} idempotents. A primitive idempotent cannot be written as $j=j_{1}+j_{2}$ for idempotents $j_{1}$ and $j_{2}$ satisfying $j_{1}j_{2}=j_{2}j_{1}=0$, except if $j_{1}$ or $j_{2}$ is zero. A primitive idempotent is characterized by the fact that $jxj=\lambda_{x}j$, $\lambda_{x}\in\mathbb C$, for any $x\in\CSn$. To a primitive idempotent is associated an irreducible representation of $\Sn$ and thus a Young tableau. The representations associated to two primitive idempotents $j$ and $j'$ are equivalent if and only if there exists $x\in\CSn$ such that $jxj'\not =0$.

A given ideal $I$ can always be decomposed as a direct sum of minimal ideals,
\be\label{Idec1} I = \bigoplus_{a}I_{a}\, .\ee
Moreover, if $I=I(j)$,
\be\label{Idec2} j = \sum_{a} j_{a}\ee
for $j_{a}\in I_{a}$, then $j_{a}j_{b}=\delta_{ab}$ and $I_{a}=I(j_{a})$. The set of irreducible representations of $\Sn$ and their multiplicities occurring in the decomposition of a given ideal $I$ is unique. Minimal ideals associated with inequivalent irreducible representations are orthogonal.

To compute the decomposition \eqref{Idec1} algorithmically, we can proceed as follows. We start from an explicit decomposition of the algebra $\CSn$,
\be\label{CSndec}\CSn = \bigoplus_{a}I(j_{a})\, .\ee
For example, we can use for the $j_{a}$ the standard Young idempotents associated with Young tableaux (note that the Young idempotents are not Hermitian in general, but this is not a problem). If we find a $j_{a_{0}}$ such that $j_{I}j_{a_{0}}\not = 0$, then the ideal $I_{a_{0}}=\{j_{I}j_{a_{0}}x\, ,\ x\in\CSn\}$ is minimal and enters into the decomposition of $I$. We compute $j_{I_{a_{0}}}$ and we iterate this process for the ideal generated by the idempotent $j_{I}-j_{I_{a_{0}}}$. This eventually yields a full decomposition of the form \eqref{Idec1}, with the additional bonus that the direct sum is automatically orthogonal.

\subsection{Tensor symmetries}

The most general notion of a tensor symmetry is described by a set of linear equations that the components of tensors having the required symmetry must satisfy. Such equations are of course invariant under arbitrary changes of basis, corresponding to $\text{GL}(d)$ transformations. For example, given a subgroup $G\subset\Sn$ and a linear character $\chi$ of $G$ (i.e.,\ a one-dimensional representation), one could consider tensors that satisfy $\sigma\cdot t = \chi(\sigma) t$ for any $\sigma\in G$. Such tensors have the ``symmetry type'' $(G,\chi)$. More generally, one can impose a set of conditions of the form
\be\label{symt} x_{i}\cdot t = 0\, ,\quad x_{i}\in\CSn\, ,\quad 1\leq i\leq p\, .\ee
Consider the ideal $I=\{x\in\CSn \mid x_{i}x=0, \ 1\leq i\leq p\}$. Then one can show that the tensors satisfying \eqref{symt} are simply the tensors of the form $x\cdot\tau$ for an arbitrary tensor $\tau$ and $x\in I$. If $I=I(j)$, these are equivalently the tensors of the form $j\cdot\tau$ or, again equivalently, the tensors satisfying
\be\label{symt2}j\cdot t = t\, .\ee
This means that the many conditions \eqref{symt} are always equivalent to the unique condition \eqref{symt2}, for a certain idempotent $j$.
For example, in the case of a tensor of symmetry type $(G,\chi)$, one has 
\be\label{Gchiex}
j_{G,\chi}=\frac{1}{|G|}\sum_{\sigma\in G}\chi(\sigma)\sigma\, ,\ee
where $|G|$ denotes the cardinal of $G$.
It is straightforward to check in this case that the unique condition $j_{G,\chi}\cdot t = t$ is equivalent to $\sigma\cdot t = \chi(\sigma) t$ for any $\sigma\in G$. The equivalence between \eqref{symt} and \eqref{symt2} remains valid in all cases. In conclusion, the \emph{symmetry types} of tensors are in one-to-one correspondence with the ideals of $\CSn$.

Let $T^{n}_{d} (I)$ denotes the vector space of rank $n$ tensors in dimension $d$ with symmetry type given by the ideal $I$. To the decomposition \eqref{Idec1} corresponds the decomposition
\be\label{TIdec} T^{n}_{d}(I) = \bigoplus_{a} T^{n}_{d}(I_{a})\, .\ee
If the ideals $I_{a}$ and $I_{a'}$ are orthogonal in $\CSn$, then $T^{n}_{d}(I_{a})$ and $T^{n}_{d}(I_{a'})$ are orthogonal in 
$T^{n}_{d}$. 
Elements of $T^{n}_{d}(I_{a})$, where $I_{a}$ is a minimal ideal, are called \emph{irreducible tensors}. Decomposing tensors into irreducible pieces can be a very useful tool which we have used to analyse the various constraints discussed in the main text.

A simple application of the above formalism is to compute the number of independent components of a tensor in $T^{n}_{d}(I)$. This is also the dimension of $T^{n}_{d}(I)$ or, from \eqref{symt2}, the rank of $\tilde\jmath$. Since $\tilde\jmath^{2}=\tilde\jmath$, $\rk\tilde\jmath = \tr\tilde\jmath$. Writing $j=\sum_{\sigma} j_{\sigma}\sigma$, the dimension can be computed by using the fact that $\tr\tilde\sigma = d^{c(\sigma)}$, where $c(\sigma)$ is the number of distinct cycles (including cycles of length one) in the cycle decomposition of $\sigma$.

\noindent\textsc{Example}: consider the Riemann tensor $R\in T^{4}_{d}$. Its symmetries are described by the equations
\be\label{Riemannsym} R_{ijkl}=-R_{jikl}=R_{klij}\, ,\quad R_{ijkl}+R_{iklj}
+R_{iljk} = 0\, .\ee
The elements $x_{i}$s in \eqref{symt} are given by
\be\label{Riemxis} x_{1}=1 + (12)\, ,\ x_{2} = 1-(13)(24)\, ,\
x_{3}= 1+(234)+(243)\, .\ee
The corresponding ideal $I_{\text{R}}$ is generated by the Young idempotent
\be\label{jRiemYoung} j_{\text{Y}_{\text{R}}} = \frac{1}{12}\bigl( 1 + (13) + (24) + (13)(24)\bigr)\bigl(1 - (12) - (34) + (12)(34)\bigr)\ee
associated with the Young tableau $\scriptsize\Yboxdim 8pt\young(13,24)$. The Young idempotent is not Hermitian, but the Hermitian generating idempotent can be found by projecting the identity element onto $I_{\text R}$,
\begin{multline}\label{jRiem}j_{I_{\text R}}=\smash{\frac{1}{24}}\bigl(
2 -2(12) + (13) + (14) + (23) + (24) -2 (34) - (123) - (124)\\ - (132)
- (134) - (142) - (143) - (234) - (243)\\ + (1234) + (1243) - 2 (1324)
+ (1342) - 2 (1423)\\ + (1432) + 2 (12)(34) + 2 (13)(24) + 2 (14)(23)\bigr)
\, .\end{multline}
A tensor $R$ has the symmetries \eqref{Riemannsym} if and only if it satisfies $j_{\text{Y}_{\text{R}}}\cdot R=R$ or equivalently $j_{I_{\text R}}\cdot R = R$. Computing the traces, one finds 
\be\label{Riemanndim}
\tr\tilde\jmath_{\text{Y}_{\text{R}}} = 
\tr\tilde\jmath_{I_{\text R}} = \frac{1}{12}d^{2}(d^{2}-1)\, ,\ee
which is the well-known number of independent components of the Riemann tensor in $d$ dimensions.

%
%

%
\subsection{A sample calculation}\label{Appsamplesec}

To illustrate the use of the above formalism on a typical example, let us give details on the derivation of the equations \eqref{decbj31}, \eqref{decbj31p} and \eqref{solbj31p} in the main text.

The first step in the calculation is to decompose the tensor $\b_{ijkl}$ into irreducible components, taking into account the constraint $\b_{ijkl}=-\b_{lkji}$. This constraint tells us that $\b_{ijkl}$ has the symmetry type of the ideal generated by the idempotent
\be\label{idemsamplegen} j=\frac{1}{2}\bigl(1-(14)(23)\bigr)\, .\ee
Using the algorithm described around equation \eqref{CSndec}, we find that
\be\label{jsampledec} j = \tilde\jmath_{3,1}+\tilde\jmath'_{3,1}+
\tilde\jmath_{2,1,1}+\tilde\jmath'_{2,1,1}\ee
with
\begin{align}\label{jsample31a}
&\tilde\jmath_{3,1}= \frac{1}{8}\bigl(1+(13)+(24)-(1234)-(1432)
-(12)(34)+(13)(24)-(14)(23)\bigr)\, ,\\
\label{jsample31b}
&\tilde\jmath'_{3,1}=\frac{1}{8}\bigl(1+(12)+(34)-(1324)-(1423)+
(12)(34)-(13)(24)-(14)(23)\bigr)\, ,
\end{align}
and similar formulas for $\tilde\jmath_{2,1,1}$ and $\tilde\jmath'_{2,1,1}$. 

In a second step, we analyse the consequences of the equation \eqref{bcond} for $n=4$,
\be\label{sampleconstraint} \partial_{i}\b_{jkl}=\bigl(J_{4}\cdot\beta
\bigr)_{ijkl}\, .\ee
By computing the decomposition of the ideal generated by $J_{4}j$, we find that it contains each Young tableau
$\smash{\Yboxdim6pt\yng(3,1)}$ and $\smash{\Yboxdim6pt\yng(2,1,1)}$ only once. This means that half of the irreducible tensors in \eqref{jsampledec} are projected out by $J_{4}$, one for each irreducible representation appearing in the decomposition, and thus only the other half will be fixed by \eqref{sampleconstraint}. 

To find out precisely which pieces are fixed by \eqref{sampleconstraint}, we proceed as follows. Let $N(J_{4})$ be the annihilating ideal associated with $J_{4}$, i.e.\
\be\label{NJ4def} N(J_{4}) = \bigl\{x\in\mathbb C[\text{S}_{4}]\mid J_{4}x=0\bigr\}\, ,\ee
and let
\be\label{intindeal} I = I(\tilde\jmath_{3,1}+\tilde\jmath'_{3,1})
\cap N(J_{4})\, .
\ee
Computing a basis for $N(J_{4})$ and then for $I$ is a simple problem of linear algebra. We can then apply our algorithm to compute the Hermitian generating idempotent $j_{3,1}$ of the ideal $I$, $I=I(j_{3,1})$, and its orthogonal $j'_{3,1}=\tilde\jmath_{3,1}+\tilde\jmath'_{3,1}-j_{3,1}$. This yields
\begin{align}\begin{split}\label{jsample31}
&j_{3,1}= \frac{1}{40}\Bigl(
5 +(12) + 4(13) +4(14) - 4(23) + 4(24) + (34) - 2(123)\\& +2(124) - 2(132)+2 (134) +2 (142) +2 (143) - 2(234) - 2(243)
- 4(1234)\\&\hskip 1.3cm -  (1324) -  (1423)
-4(1432) -3 (12)(34) + 3 (13)(24) -5 (14)(23)\Bigr) ,
\end{split}\\\label{jsample31p}
\begin{split}
&j'_{3,1}= \frac{1}{40}\Bigl(
5 +4(12) + (13)- 4(14) + 4(23) + (24) + 4(34) + 2(123)\\& -2(124)
 + 2(132)-2 (134) -2 (142) -2 (143) + 2(234) + 2(243) - (1234)
\\&\hskip 1.3cm -  4(1324) -  4(1423)
-(1432) +3 (12)(34) - 3 (13)(24) -5 (14)(23)\Bigr) .
\end{split}
\end{align}
Equations \eqref{decbj31} and \eqref{decbj31p} are obtained by acting on
the tensor $\b_{ijkl}$ with $j_{3,1}$ and $j'_{3,1}$.

By construction, $\b(j_{3,1})$ is left unconstrained by \eqref{sampleconstraint}, since $J_{4}j_{3,1}=0$. On the other hand, applying $j'_{3,1}$ to both side of \eqref{sampleconstraint} and using the properties of the primitive idempotents listed in the paragraph between equations \eqref{idealgen} and \eqref{Idec1}, we get
\be\label{intsamplecalc} j'_{3,1}\cdot\partial\b = j'_{3,1}J_{4}\,j'_{3,1}\cdot\beta = \frac{1}{2}\,j'_{3,1}\cdot\beta=\frac{1}{2}\beta(j'_{3,1})\, ,\ee
noting $\beta$ the tensor $\b_{ijkl}$ and $\partial\b$ the tensor $\partial_{i}\b_{jkl}$. This yields $\beta(j'_{3,1}) = 2j'_{3,1}\cdot
\partial\b$ which, by using the fact that $\b_{ijk}$ is expressed in terms of $\b_{[ij]}$ through equations \eqref{B3rel} and \eqref{solbj21}, finally yields the equation \eqref{solbj31p}. 

The same kind of reasoning allows to derive \eqref{decbj211}, \eqref{decbj211p} and \eqref{solbj211p} and many other results quoted in the main text.

\end{appendix}
\end{document}